\documentclass[10pt,manuscript]{aastex} %draft,manuscript,preprint, longabstract
\usepackage{latexsym,psfrag}

\newcommand{\chandra}{{\sl Chandra}}
\newcommand{\calA}{{\cal A}}

\newcommand{\bfth}{{\mathbf \theta}}

\newcommand{\rep}{^{\rm rep}}

\newcommand{\MI}{M}

\newcommand{\sherpa}{{\it Sherpa}}

\slugcomment{accepted for publication in ApJ}

\shorttitle{Calibration Uncertainty}
\shortauthors{Lee et al.}

\begin{document}

%{ABSTRACT
\title{Accounting for Calibration Uncertainties in X-ray Analysis: Effective Areas in Spectral Fitting}

\author{
Hyunsook Lee\altaffilmark{1},
Vinay L.\ Kashyap\altaffilmark{1},
David A.\ van Dyk\altaffilmark{2},
Alanna Connors\altaffilmark{3},\\
Jeremy J.\ Drake\altaffilmark{1},
Rima Izem\altaffilmark{4},
Xiao-Li Meng\altaffilmark{5}, 
Shandong Min\altaffilmark{2}, \\
Taeyoung Park\altaffilmark{6}, 
Pete Ratzlaff\altaffilmark{1}, 
Aneta Siemiginowska\altaffilmark{1}, and
Andreas Zezas\altaffilmark{7,8}
}
\affil{$^1$Smithsonian Astrophysical Observatory, 
60 Garden Street, Cambridge, MA 02138
\email{
hlee@cfa.harvard.edu \\
vkashyap@cfa.harvard.edu \\
jdrake@cfa.harvard.edu \\
rpete@head.cfa.harvard.edu \\
asiemiginowska@cfa.harvard.edu
}
}
\affil{$^2$ Department of Statistics, University of California, 
Irvine, CA 92697-1250
\email{dvd@ics.uci.edu\\
shandonm@uci.edu
}
}
\affil{$^3$Eureka Scientific, 
2452 Delmer Street Suite 100, Oakland CA 94602-3017
\email{aconnors@eurekabayes.com}
}
\affil{$^4$US Food and Drug Administration, 
Center for Drug Evaluation and Research, \\
Division of Biometrics 4,
10903 New Hampshire Ave, Silver spring, MD 20903
\email{rima.izem@fda.hhs.gov}
}
\affil{$^5$Department of Statistics, Harvard University, 
1 Oxford Street, Cambridge, MA 02138
\email{meng@stat.harvard.edu}
}
\affil{$^6$Department of Applied Statistics, Yonsei University,
Seoul 120-749, South Korea
\email{taeyoung.t.park@gmail.com}
}
\affil{$^7$ IESL, Foundation for Research and Technology, 
711 10, Heraklion, Crete, Greece
}
\affil{$^8$Physics Department, University of Crete, 
P.O. Box 2208, 710 03, Heraklion, Crete, Greece
\email{azezas@physics.uoc.edu} 
}

\begin{abstract}
While considerable advance has been made to account for statistical uncertainties in astronomical analyses, systematic instrumental uncertainties have been generally ignored.  This can be crucial to a proper interpretation of analysis results because instrumental calibration uncertainty is a form of systematic uncertainty.  Ignoring it can underestimate error bars and introduce bias into the fitted values of model parameters.  Accounting for such uncertainties currently requires extensive case-specific simulations if using existing analysis packages.  Here we present general statistical methods that incorporate calibration uncertainties into spectral analysis of high-energy data.  We first present a method based on multiple imputation that can be applied with any fitting method, but is necessarily approximate.  We then describe a more exact Bayesian approach that works in conjunction with a Markov chain Monte Carlo based fitting.  We explore methods for improving computational efficiency, and in particular detail a method of summarizing calibration uncertainties with a principal component analysis of samples of plausible calibration files.  This method is implemented using recently codified \chandra\ effective area uncertainties for low-resolution spectral analysis and is verified using both simulated and actual \chandra\ data.  Our procedure for incorporating effective area uncertainty is easily generalized to other types of calibration uncertainties.
\end{abstract}


%ABSTRACT}

%{INTRO

\section{Introduction}{\label{s:one}}

The importance of accounting for statistical errors is well established in astronomical analysis: a measurement is of little value without an estimate of its credible range.  Various strategies have been developed to compute uncertainties resulting from the convolution of photon count data with {\sl instrument calibration products} such as effective area curves, energy redistribution matrices, and point spread functions. A major component of these analyses is good knowledge of the instrument characteristics, described by the instrument calibration data.  Without the transformation from measurement signals to physically interesting units afforded by the instrument calibration, the observational results cannot be understood in a meaningful way. However, even though it is well known that the measurements of the instrument's properties (e.g., quantum efficiency of a CCD detector, point spread function of a telescope, etc.) have associated measurement uncertainties, the calibration of instruments is often taken on faith, with only nominal estimates used in data analysis, even when it is recognized that these uncertainties can cause large systematic errors in the inferred model parameters.\
\footnote{However in ground-based observations, it is customary to describe non-instrumental systematics as {\sl calibration uncertainty}, especially time-variable and foreground effects, and incorporate them in the final uncertainties. These include: e.g. atmospheric absorption effects on photometry, flat-fielding,  and astrometric calibration, as in Taris et al.\ 2011, Aguirre et al.\ 2011; calibrating brightness of distant objects in the presence of foreground dust (Conley et al 2011, Kim and Miquel 2006, Mandel et al.\ 2009). As well, uncertainties associated with the basic physics, such as e.g. specific stellar absorption lines (Thomas, Maraston, and Johansson 2010); or other model-mismatch uncertainties, such as intrinsic SN light-curve variations   (Conley et al 2011, Kim and Miquel 2006, Mandel et al.\ 2009), can also be referred to as calibration uncertainties in the literature. In this paper, we specifically concentrate on instrumental calibration uncertainties, although the formalisms introduced could in principle handle other kinds of systematic errors. }
In many subfields (exceptions include: e.g. gravitational wave astrophysics, VIRGO Collaboration 2010, LIGO Collaboration 2010 and references therein; CMB analyses, Mather et al.\ 1999, Rosset et al.\ 2010, Jarosik et al.\ 2011, and references therein; and extra-solar planet/planetary disk work, e.g. Butler et al. 1996, Maness et al. 2011, and references therein), instrument calibration uncertainty is often ignored entirely, or in some cases, it is assumed that the calibration error is uniform across an energy band or an image area.  This can lead to erroneous interpretation of the data.

Calibration products are derived by comparing data from well-defined sources obtained in strictly controlled conditions with predictions, either in the lab or using a particularly well-understood astrophysical source.  Parametrized models are fit to these data to derive best-fit parameters that are then used to derive the relevant calibration products.  The errors on these best-fit values carry information on how accurately the calibration is known and could be used to account for calibration uncertainty in model fitting. Unfortunately, however, the errors on the fitted values are routinely discarded. Even beyond the errors in these fitted values, calibration products are subject to uncertainty stemming from differences between the idealized calibration experiments and the myriad of complex settings in which the products are used. Suspected systematic uncertainty cannot be fully understood until suitable data are acquired or cross-instrument comparisons are made (David et al.\ 2007). Prospectively, this source of uncertainty is difficult to quantify but is encompassed to a certain extent in the experience of the calibration scientists.  Different mechanisms have been proposed to quantify this type of uncertainty, ranging from adopting ad hoc distributions such as truncated Gaussian (Drake et al.\ 2006) to uniform deviations over a specified range. As long as it can be characterized even loosely, statistical theory provides a mechanism by which this information can be included to better estimate the errors in the final analysis.

Users and instrument builders agree that incorporating calibration uncertainty is important (see Davis 2001; Drake et al.\ 2006; Grimm et al.\ 2009). For example, Drake et al.\ (2006) demonstrated that error bars on spectral model parameters are underestimated by as much as a factor of 5 (see their Figure~5) for high counts data when calibration uncertainty is ignored ($>>10^3$ counts for typical CCD resolution spectra). Such underestimations can lead to incorrect interpretations of the analysis results. Despite this, calibration uncertainties are rarely incorporated because only a few ad hoc techniques exist and no robust principled method is available. In short, there is no common language or standard procedure to account for calibration uncertainty.

Historically, at the International Congress of Radiology and Electricity held in Brussels in September 1910, MMe.\ Curie was asked to prepare the first standard based on high energy photon emission (X-/$\gamma$-ray): 21.99 milligrams of pure radium chloride in a sealed glass tube, equivalent to 1.67x10$^{-2}$ Curies of radioactive radium (e.g., Brown 1997 pg~9ff and references therein).  The problem then became: how to measure other samples, in reference to this standard? Although the sample preparation was done by very accurate chemistry techniques, the tricky part was designing and building the instrument to quantify the high-energy photon emission.  At the next International Committee meeting (1912, Paris) calibrating the standard was done by specialized electroscopes balancing the `ionization current' from two sources.  This instrument was deemed to have an uncertainty of one part in 400 (Rutherford and Chadwick 1911).  The original paper also describes a method for calibrating the detector. Although these measurements were quite carefully done, and complex for their time, the result was a single value (the intensity) and had a single number quantifying its error ($\frac{1}{400}$; Rutherford and Chadwick 1911).  In this case, the effect of this original unavoidable measurement error on one's final measurement of a source intensity (in Curies) is straightforward to propagate, such as by the delta-method.

Nowadays, meetings about absolute standards and measuring instruments are much more complex, incorporating multiple kinds of measurements for a single standard (e.g. CODATA; Mohr, Taylor, and Newell 2008). As well, in the general literature, one finds increasingly complex methods dealing with e.g. multivariate data and calibration (Sundberg 1999, Osbourne 1991), and even methods for `traceability' back to known standards (Cox and Harris 2006).  These approaches formulate their complexities in terms of cross-correlations of parameters.  This methodology has also been successfully used in modern astrophysics, such as in combining optical observations of supernovae for cosmological purposes (e.g. Kim and Miquel 2006).  Initially, J.~Drake and other co-authors did try formulating the dependencies and anticorrelations of the final calibration product uncertainties in terms of correlation coefficients.  However, after considerable exploration, they found this approach unable to capture the complexities of spacecraft calibration, especially at high energies. First, each part of a modern instrument such as the Chandra observatory is measured at multiple energies and multiple positions, as well as calibrating the whole system on the ground.
%[words here from   Drake??  original Cal paper??]
Second, interestingly, the instrument is modeled by a complex physics-based computer code.  The original calibration measurements are not used directly, but are benchmarks for the physical systems modeled therein.  High energy astrophysics brings a third difficulty: the previous papers assumed a Gauss-Normal distribution for the calibration-product uncertainties; this certainly does not hold for most real instruments in the high energy regime. Hence, expanding beyond Drake et al.\ (2006), in this paper, we describe how to `short-circuit' tracing back to the original calibration uncertainties by using the entire instrument-modeling code as part of statistical computing techniques.  We see this in the context of the movement towards ``uncertainty quantification'' (UQ) of large computer codes (see, e.g., Christie et al.\ 2005).

Until recently, the best available general strategy in high-energy astrophysics was to compute the root-mean-square of the measurement errors and the calibration errors and then to fit the source model using the resulting error sum (see Bevington and Robinson 1992). Unfortunately, the use and interpretation of the standard deviation relies on Gaussian errors, that the calibration errors are uncorrelated, and that the uncertainty on the calibration products can be uniquely translated to an uncertainty in each bin in data space. None of these assumptions are warranted.  Furthermore, this method, equivalent to artificially inflating the statistical uncertainty on the data, will lead to biased fits, error bars without proper coverage, and incorrect estimates of goodness of fit. Individual groups have also tried various instrument-specific methods. These range from bootstrapping (Simpson and Mayer-Hasselwander 1986) to raising and lowering response ``wings'' by hand (FOrrest 1988, Forrest Vestrand and McConnell 1997), and in one case, analytical marginalization over a particular kind of instrumental uncertainty (Bridle et al.\ 2002). In general and in important cross-instrument comparisons, however, all but the crudest methods (e.g., multiplying each instrument's total effective area by a fitted ``uncertainty factor'' as in Hanlon et al.\ 1995, Schmelz et al.\ 2009) are very difficult to handle.

Methods for handling systematic errors exist in other fields such as particle physics (Heinrich and Lyons 2007 and references therein) and observational cosmology (Bridle et al.\ 2002). In their review of systematic errors, Heinrich and Lyons (2007) advocate parameterizing the systematics into statistical models and marginalizing over the nuisance parameters of the systematics. They described various statistical strategies to incorporate systematic errors which range from simple brute force $\chi^2$ fitting to fully Bayesian hierarchical modeling. Unfortunately these analytical methods rely on Gaussian model assumption that are inappropriate for high energy astrophysics and are also highly case specific. 

Accounting for calibration uncertainty is further complicated by complex and large scale correlation in the calibration products.  The value of the calibration product at one point can depend strongly on far away values and even data collected using a different instrument. For example, the \chandra ~Low Energy Transmission Grating Spectrometer (LETGS) + High Resolution Camera - Spectroscopic readout (HRC-S) effective area is calibrated using the power-law source PKS 2155-304. Because the high-order contributions to the spectrum cannot be disentangled, the index of the power-law depends strongly on an analysis of the same source with data obtained contemporaneously with the High Energy Transmission Grating Spectrometer (HETGS) + ACIS-S. Thus, changes in the HETGS+ACIS-S effective area will affect the longer-wavelength LETGS+HRC-S effective area. The complex correlations can result in a diverse set of plausible effective area curves.  The choice among these curves can strongly affect the final best fit in day-to-day analyses. The nominally better strategy of folding the calibration uncertainty through to the final statistical errors on fitted model parameters is unfortunately unfeasible: the complex correlations make it difficult to quantify the affect on the final analysis of uncertainty in the calibration product.

Drake et al.\ (2006) proposed a strategy that accounts for these correlations by generating synthetic datasets from a nominal effective area and then fitting a model separately using each of a number of instance of a simulated effective area and then estimating the effect of the calibration error via the variance in the resulting fitted model parameters. This procedure can be implemented using standard software packages such as {\sl XSPEC} (Arnaud 1996) and \sherpa\ (Freeman et al.\ 2001, Refsdal et al.\ 2009) and demonstrates the importance of including calibration errors in data analysis. However, in practice there are some difficulties in implementing it with real data where the true parameters are not known {\sl a priori}. The ad hoc nature of the bootstrapping-type procedure means its statistical properties are not well understood, requiring the sampling distributions to be calibrated on a case-by-case basis.  That is, the procedure requires verification whenever different models are considered or different parts of the parameter space are explored. The large number of fits required also imposes a heavy computational cost.  Most importantly, it requires numerous simulated calibration products that must be supplied to end users either directly through a comprehensive database or through instrument specific software for generating them.  In general, both these strategies impose a heavy burden on calibration or analysis software maintainers.

The primary objective of this article is to propose well-defined and general methods to incorporate complex calibration uncertainty into spectral analysis in a manner that can be replicated in general practice without precise calibration expertise. Although we develop a general framework for incorporating calibration uncertainty, we limit our detailed discussion to accounting for uncertainty in the effective area for \chandra/ACIS-S in spectral analysis. We propose a Bayesian framework, where knowledge of calibration uncertainties is quantified through a prior probability.  In this way, information quantified by calibration scientists can be incorporated into a coherent statistical analysis. Operationally, this involves fitting a highly-structured statistical model that does not assume the calibration products are known fixed quantities, but rather incorporates their uncertainty through a prior distribution.  We describe two statistical strategies below for incorporating this uncertainty into the final fit. Multiple imputation fits the model several times using standard fitting routines, but with a different value of the calibration product used in each fit. Alternatively, using an iterative Markov chain Monte Carlo (MCMC) sampler allows us to incorporate calibration uncertainty directly into the fitting routine by updating the calibration products at each iteration. In either case, we advocate updating the calibration products based solely on information provided by calibration scientists and not on the data being analyzed (i.e., not updating products given the data being analyzed; see also discussion about computational feasibility in \S\ref{sec:disc:fullbayes}). This strategy leads to simplified computation and reliance on the expertise of the calibration scientists rather than on the idiosyncratic features of the data. We adopt the strategy of Drake et al.\ (2006) to quantify calibration uncertainty using an ensemble of simulated calibration products, that we call the {\sl calibration sample}. We use Principal Component Analysis (PCA) to simplify this representation.  A glossary of the terms and symbols that we use is given in Table~\ref{tab:glossary}.

In \S\ref{s:cs} we describe the calibration sample and illustrate the importance of properly accounting for calibration uncertainty in spectral analysis.  Our basic methodology is outlined in \S\ref{s:meth}, where we describe how the calibration sampler can be used to generate the replicates necessary for multiple imputation or can be incorporated into an MCMC fitting algorithm. We also show how PCA can provide a concise summary of the complex correlations of the calibration uncertainty.  Specific algorithms and strategies for implementing this general framework for spectral analysis appear in \S\ref{s:alg}. Our proposed methods are illustrated with a simulation study and an analysis of 15 radio loud quasars (Siemiginowska et al.\ 2008) in \S\ref{s:ex}.  In \S\ref{sec:disc} we discuss future directions and a general framework for handling calibration uncertainties from astrophysical observations with similar form as our yX-ray examples.  We summarize the work in \S\ref{sec:summ}.

%INTRO}

%{SAMPLE AND EFFECT

\section{The Calibration Sample and the Effect of Calibration Uncertainty}
\label{s:cs}

To coherently and conveniently incorporate calibration uncertainty into spectral fitting, we follow the suggestion of Drake et al.\ (2006) to represent it using a randomly generated set of calibration products that we call the {\sl calibration sample}. In this section we begin by describing this calibration sample, and how it can be used to represent the inherent systematic uncertainty. The methods that we discuss in this and the following sections are quite general and in principle can be applied to account for systematic uncertainty in any calibration product. For clarity, we illustrate their application to instrument effective areas.

\subsection{Representing Uncertainty}
\label{s:cs:samp}

We begin with a simple model of telescope response that assumes position and time invariance. In particular, suppose the response of a detector to an incident photon spectrum $S(E;\bfth)$,
\begin{equation} \label{eq:sim_arf}
{\cal M}(E^*;\theta)=\sum_E S(E;\bfth) A(E)P(E)R(E^*;E),
\end{equation}
where $E^*$ represents the detector channel at which a photon of energy $E$ is recorded, $\theta$ represents the parameters of the source model, and $A$, $P$, and $R$ are the effective area, point spread function, and energy redistribution matrix of the detector, respectively. We aim to develop methods to estimate $\theta$ and compute error bars that properly account for uncertainty in $A$. Of course $P$ and $R$ are also subject to uncertainty and in \S\ref{sec:disc:gen} we discuss extensions of the methods described here to handle more general sources of calibration uncertainty.

As an illustration, we consider observations obtained using the spectroscopic array of the \chandra\ AXAF CCD Imaging Spectrometer detector (ACIS-S). According to Drake et al.\ (2006), it is possible to generate a calibration sample of effective area curves for this instrument by explicitly including uncertainties in each of its subsystems (UV/Ion shield transmittance, CCD Quantum Efficiency, and the telescope mirror reflectivity). The result is a set of simulations of the effective area curves.  These encompass the range of its uncertainty, with more of the simulated curves similar to its most likely value, and fewer curves that represent possible but less likely values.  In principle, some may be more likely than others, in which case weights that indicate the relative likelihood are required.  In this article, we assume that all of the simulations in the set are equally likely, that is the simulations are representative of calibration uncertainty. The set of $L$ simulations is the {\sl calibration sample} and denoted $\calA = \{A_1, \ldots, A_L\}$, where $A_l$ is one of the simulated effective area curves.

\begin{figure}[t]
\begin{center}
\includegraphics[width=6.5in]{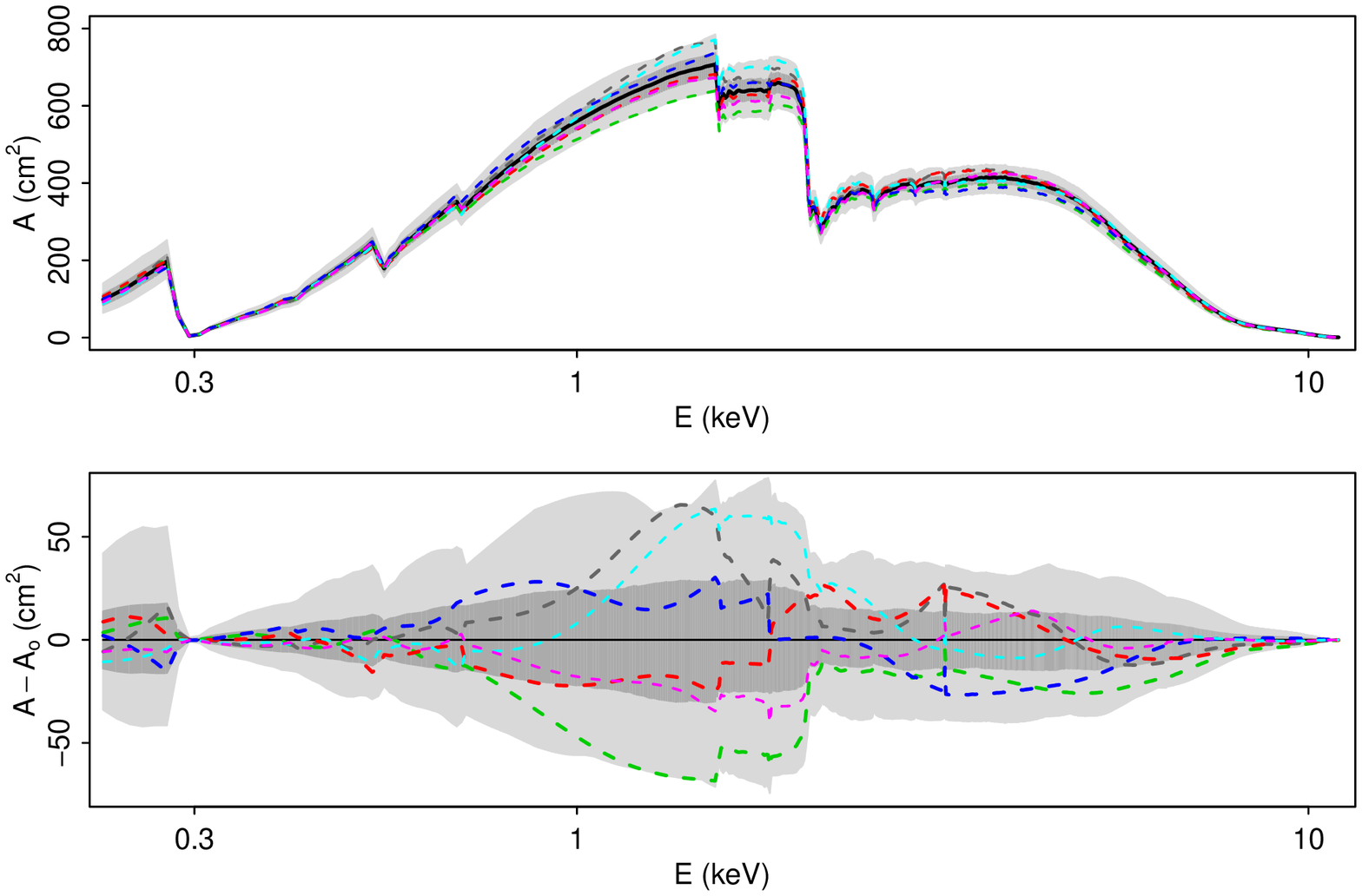}
\caption{Uncertainty in ACIS-S effective area. In the upper panel the light gray area covers all 1000 effective area curves in the calibration sample of Drake et al.\ (2006) and the darker gray area covers the middle 68\% of the curves in each energy bin. In addition six randomly selected curves are plotted as colored dashed curves and $A_0$ is plotted as a solid black curve.  The bottom panel is constructed in the same manner, but using $A_l-A_0$, in order to magnify the structure in $\calA$. The curves in $\calA$ form a complex tangle that appears to defy any systematic pattern.  As we shall see, we can use principle component analysis to form a concise summary of ${\calA}$.}
\label{fig:arf}
\end{center}
\end{figure}

The complicated structure in the uncertainty for the true effective area is illustrated in Figure~\ref{fig:arf} using the calibration sample of size $L=1000$ generated by Drake et al.\ (2006). A selection of six of the $A_l$ from $\calA$ are plotted as colored dashed lines and compared with the default effective area, $A_0$ that is plotted as a solid black line. The second panel plots the differences, $A_l - A_0$ for the same selection.  The light gray area represents the full range of $\calA$ and the dark gray area represents intervals that contain 68.3\% of the $A_l$ at each energy. The complexity of the uncertainty of $A$ is evident.  We use the calibration sample illustrated in Figure~\ref{fig:arf} as the representative example throughout this article.

\subsection{The Effect of the Uncertainty}
\label{s:cs:ex}

We discuss here the effect of the uncertainty represented by the calibration sample on fitted spectral parameters and their error bars.  We employ simulated spectra representing a broad range in parameter values.  In particular, we simulated four data sets of an absorbed power-law source with three parameters (power-law index $\Gamma$, absorption column density $N_{\rm H}$, and normalization) using the {\tt fakeit} routine in {\tt XSPECv12}. The data sets were all simulated without background contamination using the {\tt XSPEC} model {\tt wabs*powerlaw}, nominal default effective area $A_0$ from the calibration sample of Drake et al.\ (2006), and a default RMF for ACIS-S. The power law parameter ($\Gamma$), column density ($N_{\rm H}$), and nominal counts for the four simulations (see also Table~\ref{t:sim}) were
\begin{description}
\item[{\sc Simulation~1}:] $\Gamma=2$, $N_{\rm H}=10^{23}{\rm cm}^{-2}$, and $10^5$ counts;
\item[{\sc Simulation~2}:] $\Gamma=1$, $N_{\rm H}=10^{21}{\rm cm}^{-2}$, and $10^5$ counts;
\item[{\sc Simulation~3}:] $\Gamma=2$, $N_{\rm H}=10^{23}{\rm cm}^{-2}$, and $10^4$ counts; and
\item[{\sc Simulation~4}:] $\Gamma=1$, $N_{\rm H}=10^{21}{\rm cm}^{-2}$, and $10^4$ counts
\end{description}
respectively.

\begin{figure}[t]
\begin{center}
\includegraphics[width=5.0in]{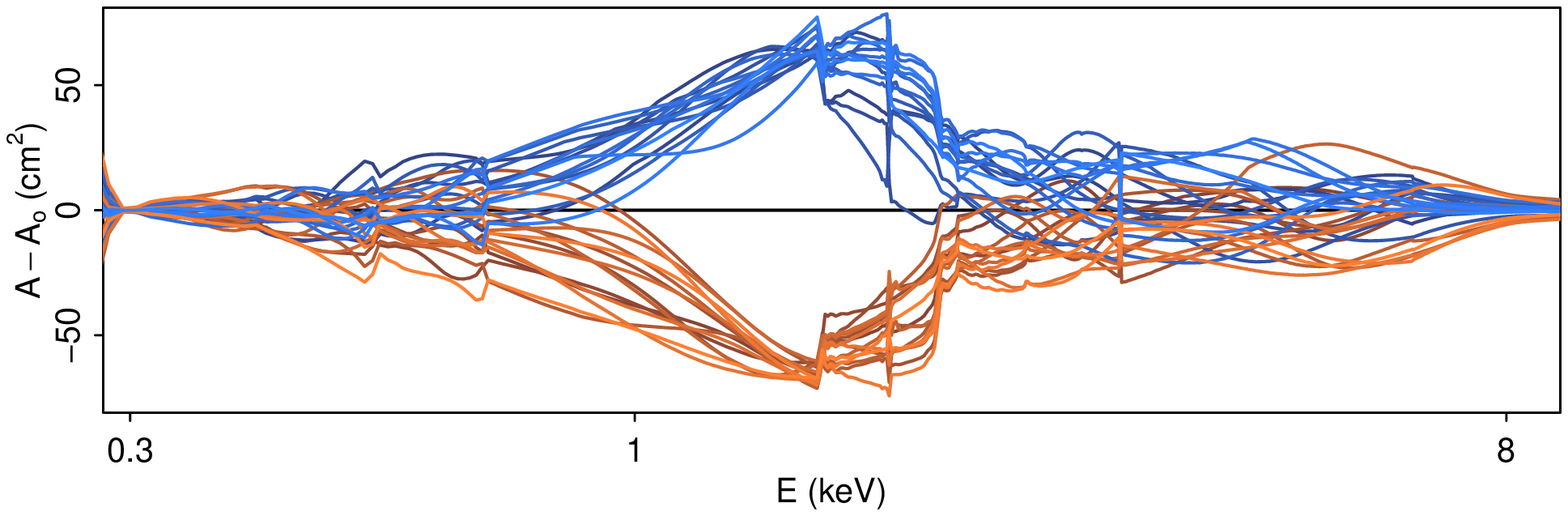}
\includegraphics[height=5.0in,angle=270]{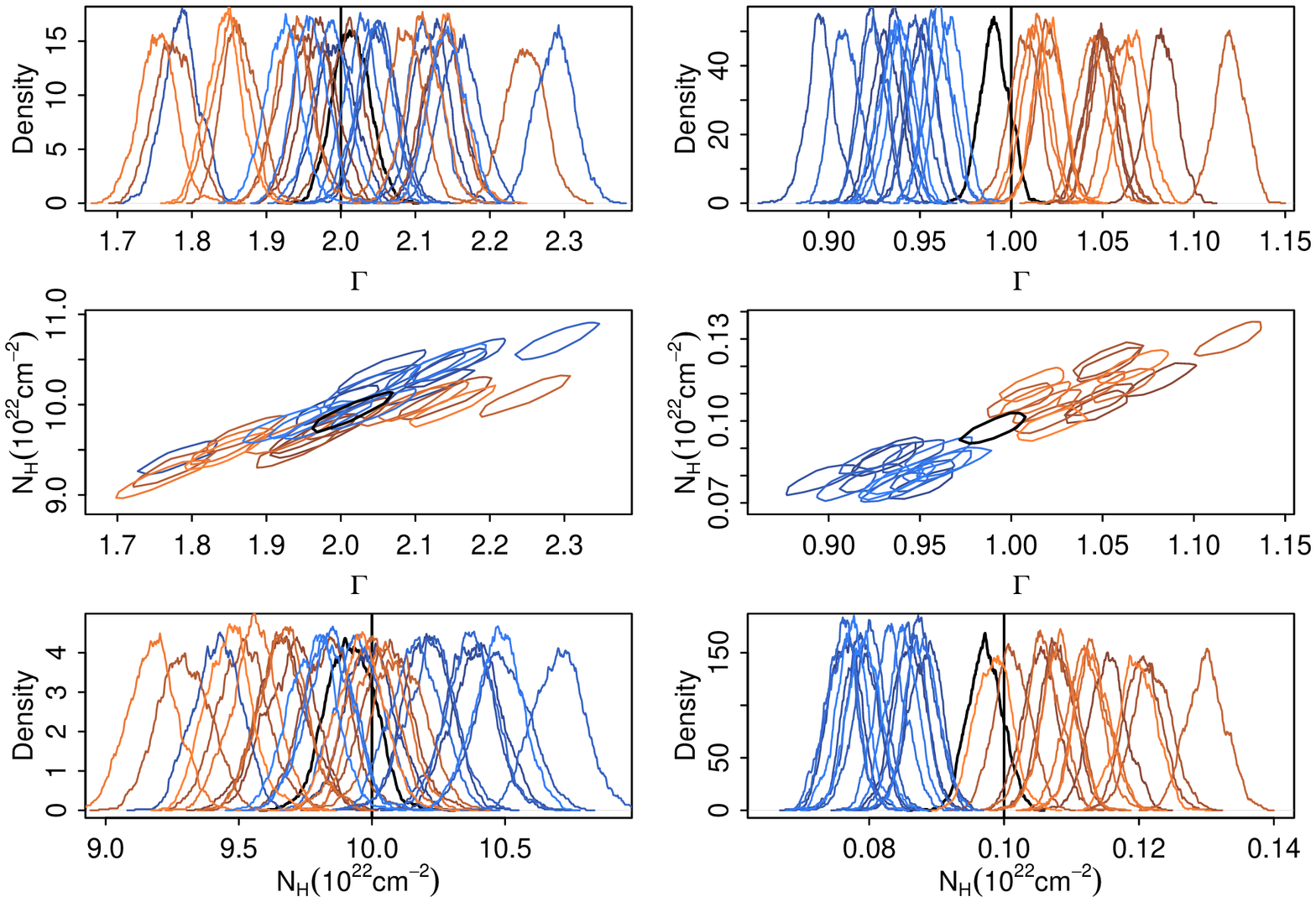}
\end{center}
\caption{\small The Effect of Calibration Uncertainty on Fitted Parameters and Error Bars.  The first panel plots the 15 effective area curves in $\calA$ with the largest  maximum in blue and the 15 curves with the smallest maximum in red, each with  $A_0$ subtracted off. The solid black horizontal line at zero represents $A_0$.  The two columns in the six lower panels correspond to {\sc Simulations~1} and 2,  respectively and plot the posterior distributions of $\Gamma$ and $N_{\rm H}$  using each of the 31 effective area curves in the first panel.  The rows of the bottom six panels correspond to  the posterior distribution of $\Gamma$, the 95.4\% contour of the joint posterior distribution, and the posterior distribution of $N_{\rm H}$.  The colors of the plotted posterior distributions indicate  the effective area curve that was used to generate the distribution.  The solid vertical black lines in the the second and fourth rows indicate  the values of the parameters used with $A_0$ to generate {\sc Simulations~1} and {\sc 2}.  The effect of the choice of effective area curves on the  posterior distributions is striking.}
\label{fig:arf_shift}
\end{figure}

To illustrate the effect of calibration uncertainty, we selected the 15 curves in $A_l\in{\calA}$ with the largest maximum values and the 15 curves with the smallest maximum values. In some sense, these are the 30 most extreme effective area curves in ${\calA}$.  They are plotted as $A_l-A_0$ in the first panel of Figure~\ref{fig:arf_shift}, along with a horizontal line at zero that represents the default ($A_0-A_0$).  We used the Bayesian method of van Dyk et al.\ (2001) to fit {\sc Simulation~1} and {\sc Simulation~2} each 31 times, using each of the 31 curves of $A_l$ plotted in Figure~\ref{fig:arf_shift}.  The resulting marginal  and joint posterior distributions for $\Gamma$ and $N_{\rm H}$ appear in rows 2-4 of Figure~\ref{fig:arf_shift}; the contours plotted in the third row correspond to a posterior probability of 95\% for each fit.\footnote{The contours in Figure~\ref{fig:arf_shift} were constructed by peeling (Green 1980) the original Monte Carlo sample. This involves removing the most extreme sampled values which are defined as the vertices of the smallest convex set containing the sample (i.e., the convex hull). This is repeated until only 95\% of the sample remains. The final hull is plotted as the contour. This is a reasonable approximation because the posterior distributions appear roughly convex.} The figure clearly shows that the effect of calibration uncertainty swamps the ordinary statistical error. The scientist who assumes that the true effective area is known to be $A_0$ may dramatically underestimate the error bars, and may miss the correct region entirely.

\begin{figure}[ht]
\begin{center}
\includegraphics[width=3.25in]{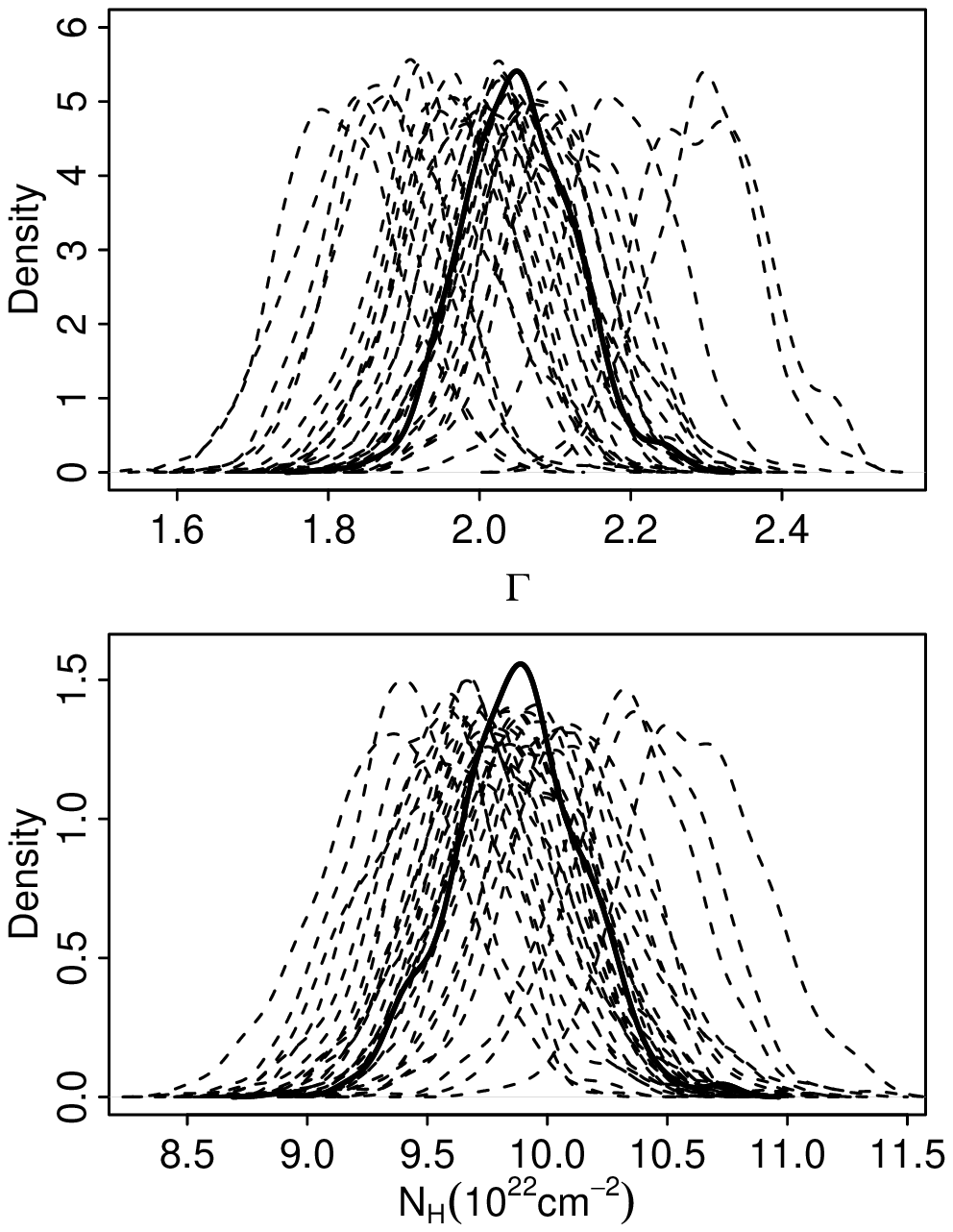}%
\includegraphics[width=3.25in]{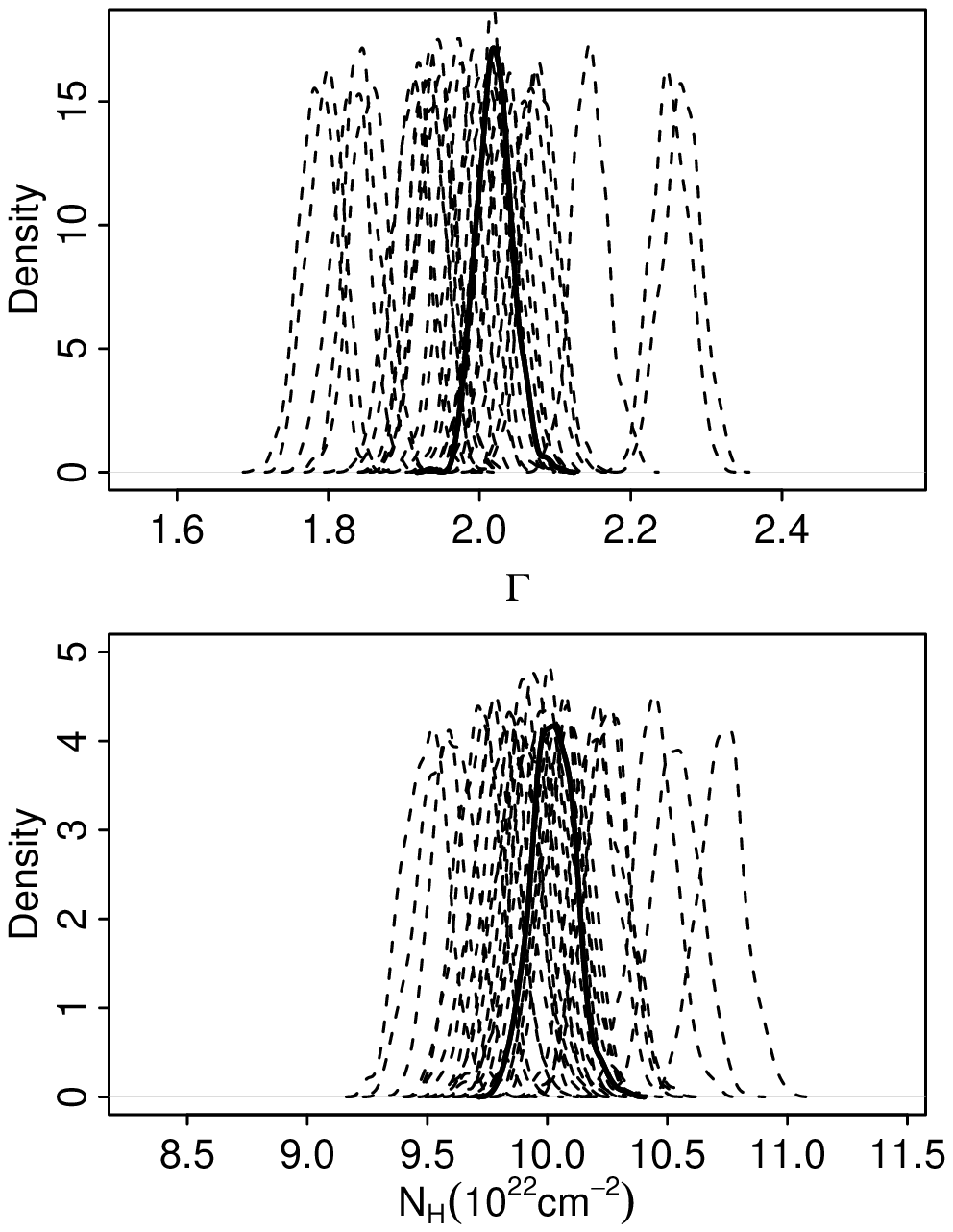}
\caption{The Interaction Between Total Counts and Calibration Uncertainty. The four panels plot the marginal posterior distributions of $\Gamma$ (row~1) and $N_{\rm H}$ (row~2) when fitting {\sc Simulation~3} (column~1 with $10^4$ counts) and {\sc Simulation~1} (column~2 with $10^5$ counts). The replicates in each panel correspond to 30 effective area curves randomly selected from ${\calA}$. The posterior distributions plotted with solid lines were constructed using $A_0$. The statistical errors are smaller with the larger data set so that calibration errors are relatively more important.
}
\label{fig:e4e5_shifts}
\end{center}
\end{figure}

As a second illustration we fit {\sc Simulation~1} and {\sc Simulation~3} each 31 times, using the same $A_l$ as in Figure~\ref{fig:arf_shift} and with $A_0$, again using the method of van Dyk et al.\ (2001). The resulting posterior distributions of $\Gamma$ and $N_{\rm H}$ are plotted in Figure~\ref{fig:e4e5_shifts}.  Comparing the two columns of the figure, the relative contribution of calibration uncertainty to the total error bars appears to grow with counts. For this reason, accounting for calibration uncertainty is especially important with rich high-count spectra. In fact, in our simulations there appears to be a limiting value where the statistical errors are negligible and the total error bars are due entirely to calibration uncertainty. The total error bars do not go below this limiting value regardless of how many counts are observed.
%if necessary, we can add a footnote here saying that in the full Bayesian case, as the sample size keeps increasing, it will become informative about the cal errors itself and eventually that too will decrease.
%In fact, there may be a limiting point where the total error bars are due entirely to calibration uncertainty and the total error bars do not go below this limiting value regardless of how many counts are observed.

We must emphasize, however, that we are assuming that the observed counts are uninformative as to which of the calibration products in the calibration sample are more or less likely. If we were not to make this assumption, however, and if a data set were so large that we were able to exclude a large portion of the calibration sample as inconsistent with the data, the remaining calibration uncertainty would be reduced and its effect would be mitigated. In this case, the default effective area and effective area curves similar to the default could potentially be found inconsistent with the data and thus the fitted model parameters could be different from what we would get if we simply relied on the default curve. In this article, however, we assume that either the data set is not large enough to be informative for the calibration products or that we do not wish to base instrumental calibration on the idiosyncrasies of a particular data set.
%In this case, if the the default effective area  and effective area curves similar to the default are found to be inconsistent with the data, the fitted model parameters may be different from what we would get if we simply relied on the default curve. In this article we do not address this situation and instead assume that either the data set is not large enough to be informative for the calibration products or that we do not wish to base instrumental calibration on the idiosyncrasies of a particular data set.

Both Figures~\ref{fig:arf_shift} and \ref{fig:e4e5_shifts} suggest that while the fitted values depend on the choice of $A\in{\cal A}$, the statistical errors for the parameters given any fixed $A\in{\calA}$ are more-or-less constant. The systematic errors due to calibration uncertainty shift the fitted value but do not effect its variance. Of course, in practice we do not know $A$ and must marginalize over it, so the total error bars are larger than any of the errors bars that are computed given a particular fixed $A$. How to coherently compute error bars that account for calibration uncertainty is our next topic.

%SAMPLE AND EFFECT}

%{ANALYSIS WITH SAMPLE
\section{Spectral Analysis Using a Calibration Sample of the Effective Area}
\label{s:meth}

In this section, we outline how the calibration sample can be used in principled statistical analyses and describe how the complex calibration sample can be summarized in a concise and complete manner using PCA.

\subsection{Statistical Analysis with a Calibration Sample}
\label{s:meth:stat}

\subsubsection{Marginalizing over Calibration Uncertainty}
\label{s:meth:stat:marg}
In a standard astronomical data analysis problem, as represented by Equation~\ref{eq:sim_arf}, it is assumed that $A\equiv{A_0}$ and that $\theta$ is estimated using $p(\theta | Y, A_0)$, where $Y$ is the observed counts and  $p$ is an objective function used for probabilistic estimation and calculation of error bars. Typical choices of $p$ are the Bayesian posterior distribution, the likelihood function, the Cash statistic, or a $\chi^2$ statistic.  We use the notation $p(\theta | Y, A_0)$ because we generally take a Bayesian perspective, with $p(\cdot)$ representing a probability distribution and the notation ``$|$'' referring to conditioning, e.g., $p(U|V)$ is to be read as ``the probability of $U$ {\sl given} that $V$ is true.''

When $A$ is unknown, it becomes a nuisance parameter\footnote{A nuisance parameter is simply an unknown but necessary parameter in the model that is not of direct interest. Its presence in the model may complicate inference, which can be a nuisance.} in the model, and the appropriate objective function becomes $p({\rm model~parameters}, A | {\rm data})$. Using Bayesian notation,
\begin{displaymath}
p(\theta, A  | Y, Z) = p(\theta | Y, Z, A) p(A|Y, Z),
\end{displaymath}
where the primary source of information for $A$ is not the observation counts, $Y$, but the large datasets and physical calcuations used by calibration scientists, and which we denote here by $Z$.
Generally speaking, we expect the information for $\theta$ to come from $Y$ rather than $Z$, at least given $A$ and we expect the information for $A$ to come from $Z$ rather than $Y$. This can be expressed mathematically by two conditional independence assumptions:
\begin{enumerate}
\item $p(\theta | Y, Z, A) = p(\theta | Y, A)$, and
\item $p(A|Y, Z) = p(A | Z)$.
\end{enumerate}
We make these conditional independence assumptions, and implicitly condition on $Z$ throughout this article. In this case, we can rewrite the above equation as
\begin{equation}
p(\theta, A  | Y) = p(\theta | Y, A) p(A),
\label{eq:marginal}
\end{equation}
which effectively replaces the posterior distribution $p(A|Y)$ with the prior distribution $p(A)$. Finally, we can focus attention on $\theta$ by marginalizing out $A$,
\begin{eqnarray}
p(\theta | Y) &\approx&  \int p(\theta | Y, A) p(A)dA \nonumber \\
&\approx& \frac1L \sum_{l=1}^L p(\theta | Y, A_l). \label{eq:newobjective}
\end{eqnarray}
That is, the objective function is simply the average of the objective functions used in the standard analysis, but with $A_0$ replaced by each of the $A_l\in\calA$. Thus, the marginalization in Equation~\ref{eq:marginal} does not necessarily involve estimating $p(A|Y)$ nor specifying a parametric prior or posterior distributions for $A$.  When this marginalization is properly computed, systematic errors  from calibration uncertainty are rigorously combined with statistical errors without need for Gaussian quadrature.

Of course, when $L$ is large as in the calibration sample of Drake et al.\ (2006), evaluating and optimizing Equation~\ref{eq:newobjective} would be a computationally expensive task. In this section we outline two strategies that aim to significantly simplify the necessary computation. The first is a general purpose but approximate strategy that can be used with any standard model fitting technique and the second is a simple adaptation that can be employed when Monte Carlo is used in Bayesian model fitting. Details and illustrations of both methods appear in \S\ref{s:alg}.

\subsubsection{Multiple Imputation}
\label{s:meth:stat:MI}
The first strategy takes advantage of a well-established statistical technique known as {\sl multiple imputation} that is designed to handle missing data (Rubin 1987, Schafer 1997).
Multiple Imputation relies on the availability of a number of Monte Carlo replications of the missing data. The replications are called the {\sl imputations} and are designed to represent the statistical uncertainty regarding the unobserved values of the missing data. Although the calibration products are not missing data {\it per se}, the calibration sample provides exactly what is needed for us to apply the method of multiple imputation: a Monte Carlo sample that represents the uncertainty in an unobserved quantity.

With the calibration sample in hand, it is straightforward to apply multiple imputation. A subset of $\calA$ of size $\MI\ll L$ is randomly selected and called the multiple imputations or the {\sl multiple imputation sample}. The standard data analysis method is then applied  $\MI$ times, once with each of the $\MI$ imputations of the calibration products. This produces $\MI$ sets of parameter estimates along with their estimated variance-covariance matrices\footnote{The variance-covariance matrix is a matrix that has the square of the error bars along its diagonal and the covariance terms as off-diagnonal elements. Recall that the covariance is the correlation times the product of the error bars: ${\rm Cov}(X, Y) = {\rm Correlation}(X,Y) \sigma_x\sigma_Y$.}, which we denote $\hat\theta_m$ and ${\rm Var}(\hat\theta_m)$, respectively, for $m=1,\ldots, \MI$. In the simplest form of the method of multiple imputation, we assume that each $\hat\theta_m$ follows a multivariate normal distribution with mean $\theta$. The final fitted values and error bars are computed using a set of simple moment calculations known as the {\sl multiple imputation combining rules} (e.g., Harel and Zhou 2005). The parameter estimate is computed simply as the average of the individual fitted values,
\begin{equation}
\hat\theta = {\frac{1}{\MI}}\sum_{m=1}^\MI \hat\theta_m.
\label{eq:mi-est}
\end{equation}
To compute the error bars, we must combine two sources of uncertainty:  the statistical uncertainty that would arise even if the calibration product were known with certainty and the systematic uncertainty stemming from uncertainty in the calibration product. Each of the $\MI$ standard analyses is computed as if the calibration product were known and therefore each ${\rm Var}(\hat\theta_m)$ is an estimate of the statistical uncertainty. Our estimate of the statistical uncertainty is simply the average of these individual estimates,
\begin{equation}
W = {\frac{1}{\MI}}\sum_{m=1}^\MI {\rm Var}(\hat\theta_m).
\label{eq:mi-win}
\end{equation}
The systematic uncertainty, on the other hand, is estimated by looking at how changing the calibration product in each of the $\MI$ analyses affects the fitted parameter. Thus, the systematic uncertainty is estimated as the variance of the fitted values,
\begin{equation}
B = {\frac{1}{\MI - 1}}\sum_{m=1}^\MI (\hat\theta_m -\hat\theta)(\hat\theta_m -\hat\theta)^\top.
\label{eq:mi-btwn}
\end{equation}
Finally the two components of variance are combined for the total uncertainty,
\begin{equation}
T = W + \left(1 + {\frac{1}{\MI}}\right) B,
\label{eq:mi-total}
\end{equation}
where the ${1\over M}$ term accounts for small number $M$ of imputations.
If $M$ is small relative to the dimension of $\theta$, $T$ will be unstable, and more sophisticated estimates should be used (e.g., Li et al.\ 1991). Here we focus on univariate summaries and error bars which depend only on one element of $\hat\theta$ and the corresponding diagonal element of $T$.

When computing the error bars for one of the univariate fitted parameters in $\hat\theta$, say component $m$ of $\hat\theta$, it is generally recommended that the number of sigma used be inflated to adjust for the typically small value of $\MI$. That is, rather than using one- and two-sigma for 68.3\% and 95.4\% intervals as is appropriate for the normal distribution, a $t$~distribution should be used, requiring a larger number of sigma to obtain 68.3\% and 95.4\% intervals. In the univariate case, the {\sl degrees of freedom} of the $t$ distribution determine the degree of inflation and can be estimated by
\begin{equation}
{\rm degrees \ of \ freedom} = (\MI-1) \left(1 + {\frac{\MI W_{mm}}{(\MI +1) B_{mm}}} \right)^2,
\label{eq:mi-df}
\end{equation}
where $W_{mm}$ and $B_{mm}$ are the $m$th diagonal terms of $W$ and $B$.

The method of multiple imputation is based on a number of assumptions.  First, it is designed to give approximate error bars on $\theta$ that include the effects of the imputed quantity, but if a full posterior distribution on $\theta$ is desired, then a more detailed Bayesian calculation must be performed (see below).  It will provide an approximately valid answer in general when the imputation model is compatible with the estimation procedure, i.e., when $\hat\theta$ is the posterior mode from essentially the same distribution as is used for the imputation (Meng 1994). Furthermore, the computed standard deviations $\sqrt{T}$ can be identified with 68\% credible intervals only when the posterior distributions are multi-variate Normal.  Additionally, when $\MI$ is small, the coverage must be adjusted using the $t$-distribution (Equation~\ref{eq:mi-df}).

\subsubsection{Monte Carlo in a Bayesian Statistical Analysis}
\label{s:meth:stat:mc}

Multiple imputation offers a simple general strategy for accounting for calibration uncertainty using standard analysis methods. Because this method is only approximate, however,  our preferred solution is a Monte Carlo method that is robust, reliable, and fast. In principle, Monte Carlo methods can handle any level of complexity present in both the astrophysical models and in the calibration uncertainty.  Monte Carlo can be used to construct powerful methods that are able to explore interesting regions in high-dimensional parameter spaces and, for instance, determine best-fit values of model parameters along with their error bars.  In this context, it is used as a fitting engine, similar to Levenberg-Marquardt, Powell, Simplex, and other minimization algorithms.  One of its main advantages is that it is highly flexible and can be applied to a wide variety of problems. A single run is sufficient to describe the variations in the model parameters that arise due to both statistical and systematic errors, which therefore leads to reduced computational costs.\footnote{In most cases, Markov chain Monte Carlo (MCMC) rather than simple Monte Carlo is required to explore complicated parameter spaces. Unfortunately, the use of MCMC in this situation raises certain technical complications. In this section we avoid these complications by focusing on the simple case of direct Monte Carlo sampling. More realistic MCMC samplers and associated complications are discussed in \S\ref{s:alg:mc}.}
Consider a Monte Carlo sample obtained by sampling the model parameters $\theta$ given the data, $Y$, and the calibration product, $A=A_0$,
$$
\theta^{(k)} \sim p(\theta|Y,A_0),
$$
where $k$ is the iteration number and $\theta^{(k)}$ are the values of the parameters at iteration $k$. The set of parameter values thus obtained is used to estimate the best-fit values and the error bars.  When calibration uncertainty is included, we can no longer condition on $A_0$ as a known value of the calibration product. Instead we add a new step that updates $A$ according to the calibration uncertainties. In particular, $\theta^{(k)}$ is updated using the same iterative algorithm as above, with an additional step at each iteration that updates $A$.  Suppose at iteration $k$, $A^{(k)}$ is the realization of the calibration product.  Then the new algorithm consists of the following two steps:
\begin{eqnarray}
\nonumber
A^{(k)} & \ {\rm is~sampled~from} \ & p(A|Y) \, \rm \ and \\
\theta^{(k)} & \ {\rm is~sampled~from} \ & p(\theta| Y, A^{(k)}).
\nonumber
\end{eqnarray}

Under the conditional independence assumptions of Section~\ref{s:meth:stat:marg}, 
we can simplify this sampler by replacing $p(A|Y)$ with $p(A)$ in the first step:
\begin{eqnarray}
A^{(k)} & \ {\rm is~sampled~from} \ & p(A) \, \rm \ and
\label{eq:mc1}\\
\theta^{(k)} & \ {\rm is~sampled~from} \ & p(\theta| Y, A^{(k)}).
\label{eq:mc2}
\end{eqnarray}
This independence assumption gives us the freedom not to estimate the posterior distribution $p(A|Y)$ and simplifies the structure of the algorithm. It effectively separates the complex problem of model fitting in the presence of calibration uncertainties into two simpler problems: (i) fitting a model with known calibration and (ii) the quantification of calibration uncertainties independent of the current data $Y$.

\subsection{Simple Summaries of a Complex Calibration Sample}
\label{s:meth:stat:pca} %TWO.3 %THREE.2?

The methods that we propose so far require storage of a large number of replicates of $A\in\calA$. Since calibration products can be observation specific this requires a massive increase in the size of calibration databases.  This concern is magnified when we consider uncertainties in the energy redistribution matrix, $R$, and point spread function, $P$, and combining multiple observations, each with their own calibration products. Although in principle this could be addressed by developing software that generates the calibration sample on the fly, we propose a more realistic and immediate solution that involves statistical compression of $\calA$. Compression of this sort takes advantage of the fact that many of the replicates in $\calA$ differ very little from each other and in principle we can reduce the sample's dimensionality from thousands to only a few with little loss of information. Here we describe how principal component analysis (PCA) can accomplish this for the  {\sl Chandra}/ACIS-S calibration sample generated by Drake et al.\ (2006) and illustrated in Figure~\ref{fig:arf}.

PCA is  a commonly applied linear technique  for dimensionality reduction and data compression (Jolliffe 2002, Anderson 2003, Ramsay and Silverman 2005, Bishop 2007). Mathematically, PCA is defined as an orthogonal linear transformation of a set of variables such that the first transformed variable defines the linear function of the data with the greatest variance, the second transformed variable define the linear function {\it orthogonal to the first} with the greatest variance, and so on. PCA aims to describe variability and is generally computed on data with mean zero. In practice, the mean of the data is subtracted off before the PCA and added back after the analysis. Computation of the orthogonal linear transformation is accomplished with the singular value decomposition of a data matrix with each variable having mean zero. This generates a set of eigenvectors that correspond to the orthogonal transformed variables, along with their eigenvalues that indicate the proportion of the variance correlated with each eigenvector. The eigenvectors with the largest values are known as the {\sl principal components}. By selecting a small number of the largest principal components, PCA allows us to effectively summarize the variability of a large data set with a handful of orthogonal eigenvectors and their corresponding eigenvalues.

Our aim is to effectively compress $\calA$ using PCA. Using the singular vector decomposition of a matrix with rows equal to the $A_l-\bar A$ with $\bar A ={\frac{1}{L}} \sum_l A_l$, we compute the eigenvectors $(v_1, \ldots v_L)$ and corresponding eigenvalues $(r_1^2, \ldots r_L^2)$, ordered such that $r_1 \geq r_2 \geq ... \geq r_{L}.$  The fraction of the variance of $\calA$ in the direction of $v_l$ is
\begin{equation}
f_l= \frac{r_l^2}{{\sum_{j=1}^{L}} r_j^2}.
\label{eq:pcafrac}
\end{equation}
In practice, this gives us the option of using a smaller number of components, $J<L$ in the reconstruction, that is sufficient to account for a certain fraction of the total variance. A large amount of compression can be achieved because very few components are needed to compute the effective area to high precision. For example, in the case of ACIS effective areas, 8-10 components (out of 1000) can account for 95\% of the variance, and $\approx20$ components can account for 99\% of the variance. Note that this approximation is valid only when considered over the full energy range; small localized variations in $\calA$ that contribute little to the total variance, even if they may play a significant role in specific analyses (the depth of the C-edge, for example) may not be accounted for.

With the PCA representation of $\calA$ in hand, we wish to generate replicates of $A$ that mimic $\calA$. In doing so, however, we must account for the fact that calibration products typically vary from observation to observation to reflect deterioration of the telescope over time and other factors that vary among the observations.  However, even though the magnitudes of the calibration products may change, the underlying uncertainties are less variant and are comparable across different regions of the detector at different times. We thus suppose that the differences among the calibration samples can be represented by simply changing the default calibration product, at least in many cases.  That is, we assume that the distribution in the calibration samples differ only in their (loosely defined) average and that differences in their variances can be ignored. Under this assumption, we can easily generate calibration replicates based on the first $J$ principal components as
\begin{eqnarray}
A\rep & = & \bar A +  (A^*_0 - A_0) + \sum_{j=1}^{J} e_j r_j v_{j} +\xi e_{J+1},
\label{eq:pca0} \\
           &=  & A^*_0 +\delta\bar A + \sum_{j=1}^{J} e_j r_j v_{j} +\xi e_{J+1}
\label{eq:pca}
\end{eqnarray}
where $A^*_0$ is the observation-specific effective area that would currently be created by users, $A_0$ is the nominal default effective area from calibration, $\delta\bar A = \bar A - A_0$, $\xi = \sum_{j=J+1}^L r_j v_{j}$, and $(e_1, \ldots, e_{J+1})$  are independent standard normal random variables. In addition to the first $J$ principal components, this representation aims to improve the replicates by including the residual sum of the remaining $L-J$ components.  Equation~\ref{eq:pca0} shows how we account for $A^*_0$. If $A^*_0$ were equal to $A_0$, Equation~\ref{eq:pca0} would reduce to the standard PCA representation. To account for the observation-specific effective area, we add the offset $A^*_0 - A_0$.  Equation~\ref{eq:pca} rearranges the terms to express $A\rep$ as the sum of calibration quantities that we propose to provide in place of $\calA$. In particular, using Equation~\ref{eq:pca}, we can generate any number of Monte Carlo replicates from $\calA$, using only $\delta\bar A$, $A^*_0$, $(r_1 v_1, \ldots, r_Lv_L)$, and $\xi$.  In this way we need only provide instrument-specific and not observation-specific values of $(r_1 v_1, \ldots, r_Lv_L)$, and $\xi$.

\begin{figure}[t]
\begin{center}
\includegraphics[width=6.5in]{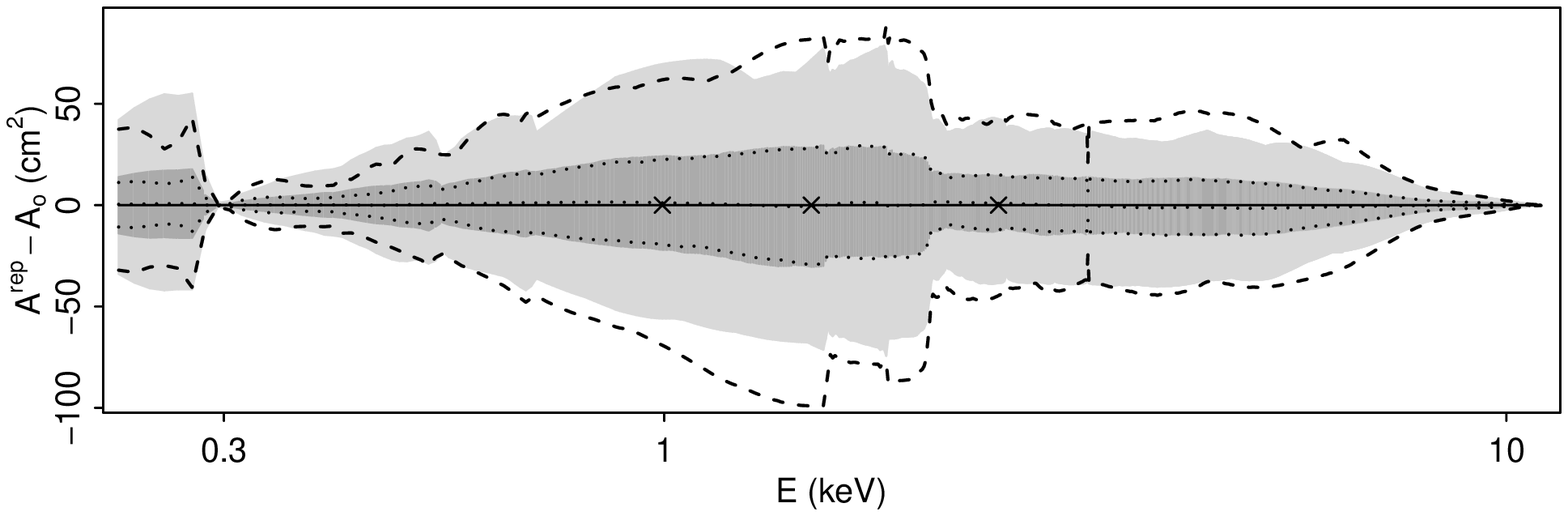}
\includegraphics[width=6.5in]{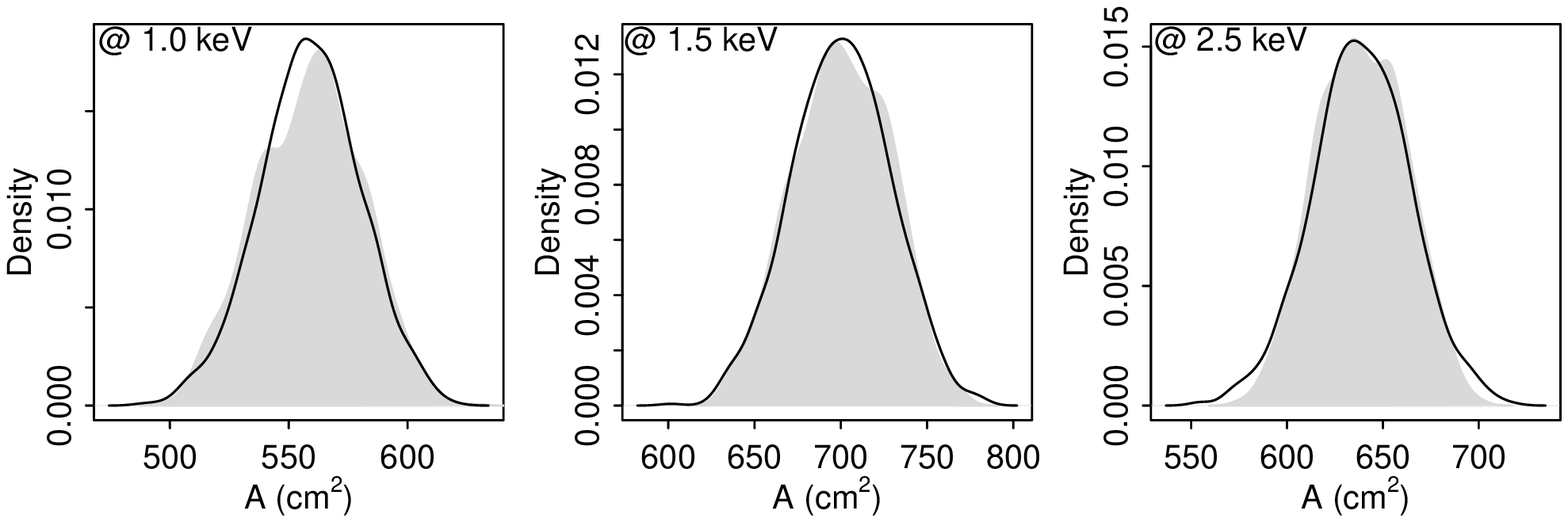}
\caption{Summarizing the Calibration Sample Using PCA. The grey regions  in the upper left panel are identical to those in the second panel of Figure~\ref{fig:arf} and give intervals for each energy bin that contain 100\% and 68.3\% of the calibration sample.  The dashed and dotted lines outline  intervals for each energy bin containing 100\% and 68.3\% of 1000 PCA replicates  of the effective area, sampled using Equation~{\ref{eq:pca0}. The correspondence  between the calibration sample and the PCA sample is quite good, especially for the 68.3\% intervals. The solid horizontal line is $A_0$ and dotted  line near it is the almost identical $\bar A$. The other three panels give  histograms of the calibration sample (grey) and the PCA sample (solid line)  in each of three energy bins, represented by $\times$ signs in the first panel.}}
\label{fig:pca}
\end{center}
\end{figure}

Figure~\ref{fig:pca} illustrates the use of PCA compression on the calibration sample generated by Drake et al.\ (2006) and illustrated in  Figure~\ref{fig:arf}. We generated 1000 replicate effective areas using Equation~\ref{eq:pca} with $A_0=A_0^*$ and  $J=8$. The dashed and dotted lines in the upper left panel respectively superimpose the full range and 68.3\% intervals of these replicates on the corresponding intervals for the origi
nal calibration sample, plotted in light and dark grey.  In this case, using $J=8$ captures  96\% of the variation in $\calA$, as computed with Equation~\ref{eq:pcafrac}. The remaining three panels give cross sections at 1.0, 1.5, and 2.5 keV. The distributions of the 1000 replicates generated using Equation~\ref{eq:pca} appears as solid lines, and those of the original calibration sample as a gray regions. The figure shows that PCA replicates generated with $J=8$ are quite similar to the original calibration sample.

Although the PCA representation cannot be perfect (e.g., it does not fully represent uncertainty overall or in certain energy regions) it is much better than not accounting for uncertainty at all.

%ANALYSIS WITH SAMPLE}

%{ALGORITHMS
\section{Algorithms Accounting for Calibration Uncertainty}
\label{s:alg}

In this section we describe specific algorithms that incorporate calibration uncertainty into standard data analysis routines. In \S\ref{s:alg:mi} we show how multiple imputation can be used with  popular scripted languages like {\sl HEASARC}/XSPEC and {\sl CIAO/\sherpa}\ for spectral fitting, and in \S\ref{s:alg:mc} we describe some minor changes that can be made to sophisticated Markov chain Monte Carlo samplers to include the calibration sample. In both sections we begin with cumbersome but precise algorithms and then show how approximations can be made to simplify the implementation. Our recommended algorithms appear in \S\ref{s:alg:mi:pca} and \S\ref{s:alg:prag:pca}. In \S\ref{s:ex} we demonstrate that these approximations have a negligible effect on the final fitted values and error bars.

\subsection{Algorithms for Multiple Imputation}
\label{s:alg:mi}
\subsubsection{Using the Full Calibration Sample}
\label{s:alg:mi:full}
Multiple imputation is an easy to implement method that relies heavily on standard fitting routines. An algorithm for accounting for calibration uncertainty using multiple imputation can is described by:

\begin{description}
\item[\sc Step~1:] For $m=1,\ldots, \MI$, repeat the following:
\begin{description}
\item[\sc Step~1a:]  Randomly sample $A_m$ from ${\calA}$.
\item[\sc Step~1b:]  Fit the spectral model (e.g., using \sherpa) in the usual way, but with effective area set to $A_m$
\item[\sc Step~1c:]  Record the fitted values of the parameters as $\hat\theta_m$
\item[\sc Step~1d:]  Compute the variance-covariance matrix of the fitted values and record the matrix as ${\rm Var}(\hat\theta_m)$. (In \sherpa\ this can be done using the {\tt covariance} function.)
\end{description}
\item[\sc Step~2:]  Use Equation~\ref{eq:mi-est} to compute the fitted value, $\hat\theta$ of $\theta$.
\item[\sc Step~3:]  Use Equations~\ref{eq:mi-win}--\ref{eq:mi-total} to compute the variance-covariance matrix, ${\rm Var}(\hat\theta) = T$, of $\hat\theta$. The square root of the diagonal terms of ${\rm Var}(\hat\theta) = T$ are the error bars of individual parameters.
\item[\sc Step~4:] Use Equation~\ref{eq:mi-df} to compute the degrees of freedom for each component of $\hat\theta$ which are used to properly calibrate the error bars computed in {\sc Step}~3.
\end{description}
Asymptotically, $\pm{1}\sigma$ error~bars correspond to equal-tail $68.3\%$ intervals under the Gaussian distribution. When the number of imputations is small, $\pm{t_{\rm df}}\sigma$ error~bars should be used instead, where $t_{\rm df}$, a number $>1$, can be looked up in any standard $t$-distribution table using ``df'' equal to the degrees of freedom computed in {\sc Step~4}, see \S\ref{s:ex:mi} for an illustration.

If the correlations among the fitted parameters are not needed, the error bars of the individual fitted parameters can be computed one at a time using Equations~\ref{eq:mi-win}--\ref{eq:mi-total} with $\hat\theta_m$ and ${\rm Var}(\hat\theta_m)$ replaced by the fitted value of the individual parameter and the square of its error bars, both computed using $A_m$.

\subsubsection{Using the PCA Approximation}
\label{s:alg:mi:pca}
Using the PCA approximation results in a simple change to the algorithm in \S\ref{s:alg:mi:pca}: {\sc Step~1a} is replaced by (see Equation~\ref{eq:pca}):
\begin{description}
\item[\sc Step~1a:]  Set $A_m =  A^*_0  +  \delta \bar A + \sum_{j=1}^{J} e_j r_j v_{j} +\xi e_{J+1}$, where $(e_i,\ldots, e_{J+1})$ are independent standard normal random variables.
\end{description}
The choice between this algorithm and the one described in Section~\ref{s:alg:mi:full} should be determined by the availability of a sample of size $M$ from {\cal A} (in which case the Algorithm in Section~\ref{s:alg:mi:full} should be used) or of the PCA summaries of {\cal A} required for the algorithm in this section.

\subsection{Algorithms for Monte Carlo in a Bayesian Analysis}
\label{s:alg:mc}

In \S\ref{s:meth:stat:mc} we considered simple Monte Carlo methods that simulate directly from the posterior distribution, $\theta^{(k)} \sim p(\theta | Y, A_0)$. More generally, Markov chain Monte Carlo (MCMC) methods can be used to fit much more complicated models. (Good introductory references to MCMC can be found in Gelman 2003 and Gregory 2005.)
A Markov chain is an ordered sequence of parameter values such that any particular value in the sequence depends on the history of the sequence only through its immediate predecessor. In this way MCMC samplers produce dependent draws from $p(\theta|Y,A_0)$ by simulating $\theta^{(k)}$ from a distribution that depends on the previous value of $\theta$ in the Markov chain, $\theta^{(k)}~\sim{\cal~K}(\theta|\theta^{(k-1)};Y, A_0)$.  That is, ${\cal K}$ is designed to be simple to sample from, while the full $p(\theta | Y, A_0)$ may be quite complex. The price of this, however, is that the $\theta^{(k)}$ may not be statistically independent of the $\theta^{(k-1)}$; and in fact may have appreciable correlation with $\theta^{(k-d)}$ (that is, an autocorrelation of length $d$).
The distribution ${\cal K}$ is derived using methods such as the Metropolis-Hastings algorithm and/or the Gibbs sampler that ensures that the resulting Markov chain converges properly to $p(\theta | Y, A_0)$. Van Dyk et al.\ (2001) show how Gibbs sampling can be used to derive ${\cal K}$ in high-energy spectral analysis. Their method has recently been generalized in a \sherpa\ module called {\tt pyBLoCXS} (Bayesian Low Count X-ray Spectral analysis in Python, to be released)\footnote{URL: {\tt http://cxc.harvard.edu/sherpa}. The {\tt pyBLoCXS} routine uses a different choice of ${\cal K}$ that relies more heavily on Metropolis-Hastings than on Gibbs sampling and can accommodate  a larger class of spectral models.} In this section we show how {\tt pyBLoCXS} can be modified to account for calibration uncertainty. For clarity we use the notation
\begin{equation}
\theta^{(k)}\sim {\cal K}_{\tt pyB} (\theta | \theta^{(k-1)}; Y, A)
\label{eq:pyB}
\end{equation}
 to indicate a single iteration of {\tt pyBLoCXS} run with the effective area set to $A$.

\subsubsection{A Pragmatic Bayesian Method}
\label{s:alg:prag}

In \S\ref{s:meth:stat:mc} we describe how a Monte Carlo sampler can be constructed to account for calibration uncertainly under the assumption that the observed counts carry little information as to the choice of effective area curve. In particular, we must iteratively update $A^{(k)}$ and $\theta^{(k)}$ by sampling them as described in Equations~\ref{eq:mc1} and \ref{eq:mc2}.   Sampling $A^{(k)}$ from $p(A)$  can be accomplished by simply selecting an effective area curve at random from $\calA$.  Updating $\theta$ is more complicated, however, because we are using MCMC. We cannot directly sample $\theta^{(k)}$ from $p(\theta | Y, A^{(k)})$  as stipulated by Equation~\ref{eq:mc2}. The {\tt pyBLoCXS} update of $\theta^{(k)}$ depends on the previous iterate, $\theta^{(k-1)}$. Thus, we must iterate {\sc Step~2} of the fully Bayesian sampler several times before it converges and delivers an uncorrelated draw from $p(\theta | Y, A^{(k)})$.  In this way, we iterate {\sc Step~2}  in the following sampler until the dependence on $\theta^{(k-1)}$ is negligible. To simplify notation, we display iteration $k+1$ rather than iteration $k$; notice that after $I$ repetitions, {\sc Step~2} returns $\theta^{(k+1)}$.  In practice we expect a relatively small value of $I$ ($\sim 10$ or fewer) will be sufficient, see \S\ref{s:ex:pBayes}. The MCMC step for a given $k$ is as follows:
\noindent
\begin{description}
\item[\sc Step~1:] Sample $A^{(k+1)} \sim p(A)$.
\item[\sc Step~2:] For $i=1,\ldots, I$,
\item[\quad\quad\quad \,] Sample $\theta^{(k + i/I)}\sim {\cal K}_{\tt pyB} (\theta | \theta^{(k+ (i-1)/I)}; Y, A^{(k+1)})$.
\end{description}
Once the MCMC sampler run is completed, the `best-fit' and confidence bounds for each parameter are typically determined from the mean and widths of the histograms constructed from the traces of $\{\theta^{k}\}$; or mean and widths of the contours (for multiple parameters), as in Figures~\ref{fig:arf_shift} and \ref{fig:comp_e5_1to4}; see Park et al.\ (2008) for discussion.

%Because we assume that $p(A|Y) \approx p(A)$, that is, the current observation carries little information for the calibration product, we may replace {\sc Step~1} of the fully Bayesian method with: Sample $A^{(k)} \sim p(A|\theta^{(k-1)})$. Because we are not assuming an {\it a priori} relationship between $\theta$ and $A$, $p(A|\theta) = p(A)$.  Thus sampling $A^{(k)} \sim p(A|\theta^{(k-1)})$ is equivalent to: Sample $A^{(k)} \sim p(A)$. Even if $p(A|Y) = p(A)$, however, $p(A|\theta, Y)$ may not be equal to $p(A|\theta)$. In fact we expect $p(A|\theta, Y)$ to depend heavily on $\theta$ in some cases, see \S\ref{s:alg:mc}.1. This is true even we assume $p(A|\theta) =p(A)$, {\it a priori}. Thus, the substitution of $A^{(k)} \sim p(A)$ for {\sc Step~1} may not be strictly valid, see Appendix~\ref{s:app}. To address this, we consider the psudo algorithm given in Equation~\ref{eq:mc}. Because this is a simple Monte Carlo sampler, the first step samples $A^{(k)}$ from $p(A|Y)\approx p(A)$ and avoid the complication caused by conditioning on $\theta^{(k-1)}$. The challenge is that the second step must produce a direct sample of $p(\theta|Y,A^{(k)})$ that does not depend on $\theta^{(k-1)}$. We can accomplish this by iterating {\sc Step~2} until the dependence $\theta^{(k-1)}$ is negligible. To simplify notation in the following algorithm, we display iteration $k+1$ rather than iteration $k$; notice that after $I$ repetitions, {\sc Step~2} returns $\theta^{(k+1)}$.

\subsubsection{A Pragmatic Bayesian Method with the PCA Approximation}
\label{s:alg:prag:pca}

Using the PCA approximation results in a simple change to the algorithm in \S\ref{s:alg:prag}: {\sc Step~1} is replaced by
\begin{description}
\item[\sc Step~1:]  Set $A^{(k+1)} =  \delta \bar A +  A^*_0 + \sum_{j=1}^{J} e_j r_j v_{j} +\xi e_{J+1}$, where $(e_1,\ldots, e_{J+1})$ are independent standard normal random variables.
\end{description}
Because of the advantages in storage that this method confers, and the negligible effect that the approximation has on the result (see \S\ref{s:ex:comp}), this is our recommended method when using MCMC to account for calibration uncertainty with data sets with ordinary counts.

%ALGORITHMS}

%{NUMERICAL EVALUATION
\section{Numerical Evaluation}
\label{s:ex}

In this section we investigate optimal values of the tuning parameters needed by the algorithms and compare the performance of the algorithms with simulated and with real data. Throughout, we use the absorbed power law simulations described in Table~\ref{t:sim}  to illustrate our methods and algorithms.  The eight simulations represent a $2\times 2\times 2$  design with the three factors being (1) data simulated with $A_0$ and with an extreme effective area curve from $\calA$, (2) $10^5$ and $10^4$ nominal counts, and (3) two power law spectral models. These simulations include the four described in \S\ref{s:cs:ex}.  We investigate the number of imputations required in Multiple Imputation studies in \S\ref{s:ex:mi}, and the number of subiterations required in MCMC runs in \S\ref{s:ex:pBayes}.  We compare the results from the different algorithms (Multiple Imputation with sampling and with PCA, and {\tt pyBLoCXS} with sampling and PCA) in detail in \S\ref{s:ex:comp}, and apply them to a set of Quasar spectra in \S\ref{s:ex:quasar}.

\subsection{Determining the Number of Imputations}
\label{s:ex:mi}

\begin{figure}[t]
\begin{center}
\psfrag{l}{$\MI$}
\includegraphics[height=5.0in,angle=270]{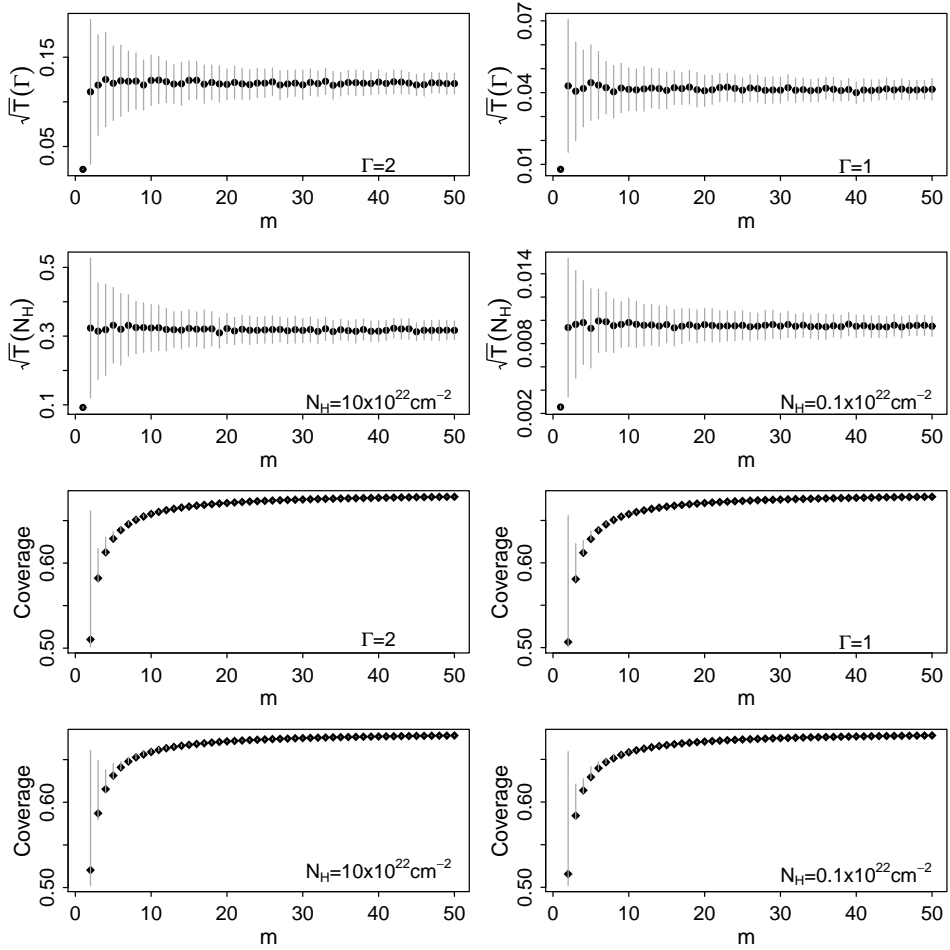}
\caption{
\baselineskip=15pt
The Sensitivity of the Error Estimates on the Number of Imputations, $\MI$.  We show the result of varying $\MI$ on fits carried out for spectra from {\sc Simulation~1} (left column) and {\sc Simulation~2} (right column).  For each $\MI=m$, we generate $m$ effective area curves $\{A\rep_i, i=1,\ldots,m\}$ using Equation~\ref{eq:pca}, and carry out separate fits for each using \sherpa, and combine the the results of the fits using the multiple imputation combining rules (Equations~\ref{eq:mi-est}--\ref{eq:mi-total}).  This gives us one value for the combined (statistical and systematic) error bar.  We repeat this process 200 times for each $m$ to investigate the variability of the computed error bar.  The average computed errors (filled symbols) are shown for the power-law index $\Gamma$ (top row) and the absorption column density $N_{\rm H}$ (second row) as a function of $m$ along with the uncertainty on the errors due to sampling (thin vertical bars).  The total error is grossly underestimated for $m=1$ (computed for only the default effective area), and the uncertainty on the error decreases for $m>1$.  Typically, $\MI\approx{20}$ is sufficient to obtain a reasonably accurate estimate of the total error.  We also show the coverage fraction for the derived error bars for $\Gamma$ (third row from the top) and $N_{\rm H}$ (bottom row).  The coverage is small for small $m$ because the degrees of freedom is small (see Equation~\ref{eq:mi-df}) but asymptotically approaches Gaussian coverage of $0.683$ for large $m$.
}
\label{fig:mi}
\end{center}
\end{figure}

When using multiple imputation, we must decide how many imputations are required to adequately represent the variability in ${\calA}$. Although in the social sciences as few as 3-10 imputations are sometimes recommended (e.g., Schafer 1997), larger numbers more accurately represent uncertainty. To investigate this we fit spectra from {\sc Simulation~1} and {\sc Simulation~2} using {\it Sherpa}, with different values of $\MI$, the number of imputations. For each value of $\MI$ we generate $\MI$ effective area curves, $A\rep$, using Equation~\ref{eq:pca}, fit the simulated spectrum $\MI$ times, once with each $A\rep$, derive the $1\sigma$ error bars, and combine the $\MI$ fits using the multiple imputation combining rules in Equations~\ref{eq:mi-est}--\ref{eq:mi-total}. This gives us a single total error bar for each parameter. We repeat this process 200 times for each value of $\MI$ to investigate the variability of the computed error bar for each value of $\MI$. The result appears in the first two rows of Figure~\ref{fig:mi}.  For small values of $\MI$ the error bars are often too small or too large. With $\MI$ larger than about 20, however, the multiple imputation error bars are quite accurate.  Even with $\MI=2$, however, the error bars computed with multiple imputation are more representative of the actual uncertainty than when we fix the effective area at $A_0$, which is represented by $\MI=1$ in Figure~\ref{fig:mi}. Generally speaking, $\MI=20$ is usually adequate, but $\MI=20$ to $50$  is better if computational time is not an issue.
Note that the size of the calibration sample ${\calA}$ is generally much larger than this, and it is therefore a fair sample to use in the Bayesian sampling techniques described in \S\ref{s:alg:mc}.

When $\MI$ is relatively small, the computed $\pm{1}\sigma$ error bars may severely underestimate the uncertainty, and must be corrected for the degrees of freedom in the imputations (see Equation~\ref{eq:mi-df}).  To illustrate this, we compute the nominal coverage of the standard  $\pm{\rm one}\sqrt(T)$ interval for each of the MI analyses described in the previous paragraph. When $\MI$ is large, such intervals are expected to contain the true parameter value 68.3\% of the time, the probability that a Gaussian random variable is within one standard deviation of its mean. With small $\MI$, however, the coverage decreases because of the extra uncertainty in the error bars. The bottom two rows of Figure~\ref{fig:mi} illustrate the importance of adjusting for the degrees of freedom, especially when using relatively small values of $M$. The plots give the range of nominal coverage rates for one $\sqrt{T}$ error bars. For large $\MI$ the coverage approaches $95\%$, but for small $\MI$ it can be as low as 50-60\%. This can be corrected by computing the degrees of freedom using Equation~\ref{eq:mi-df} and using $\pm{t_{\rm df}}\sigma$ instead of $\pm{\rm one }\sqrt{T}$, as described in \S\ref{s:alg:mi:full}. 
%{\bf [XLM and DvD ask: why not include this in the simulation results?] [this will eventually go into the figure, viz. Hyunsook's R code that has been sent to DvD; but let us not wait upon this modification before submission]}

\subsection{Determining the Number of Subiterations in the Pragmatic Bayesian Method}
\label{s:ex:pBayes}

%As noted in \S\ref{s:alg:prag}, in order to obtain a sample from the $\theta^{(t)}\sim p(\theta|Y, A^{(t)})$ as in Equation~\ref{eq:mc2} we must iterate {\tt pyBLoCXS} $I$ times to eliminate the dependence of $\theta^{(k-1)}$. To investigate how large $I$ must be, we run {\tt pyBLoCXS} on the spectra from {\sc Simulations~1} and {\sc Simulation~5} of Table~\ref{t:sim}, which were generated using the ``default'' and an ``extreme'' effective area curve. Since {\sc Simulation~5} was generated using the ``extreme'' effective area curve, it is the ``extreme'' curve that is actually ``correct'' and the ``default'' curve that is ``extreme''. The resulting autocorrelation plots (and parameter traces) for $\Gamma$ appear in Figure~\ref{fig:acf}.  These plots report the correlation of $\theta^{(k)}$ and $\theta^{(k+I)}$ for each value of $I$ plotted on the horizontal axis. The plots show that for $I>10$ the autocorrelations are essentially zero for both spectra, and we can consider $\theta^{(k)}$ and $\theta^{(k+10)}$ to be independent. Similar plots for $N_{\rm H}$ and the normalization parameter (not included) are essentially identical. Thus, in all subsequent computations we set $I=10$ in the pragmatic Bayesian samplers. Generally speaking, the user should construct autocorrelation plots to determine how large $I$ must be in a particular setting.

\begin{figure}[t]
\begin{center}
\includegraphics[width=6.5in,angle=0]{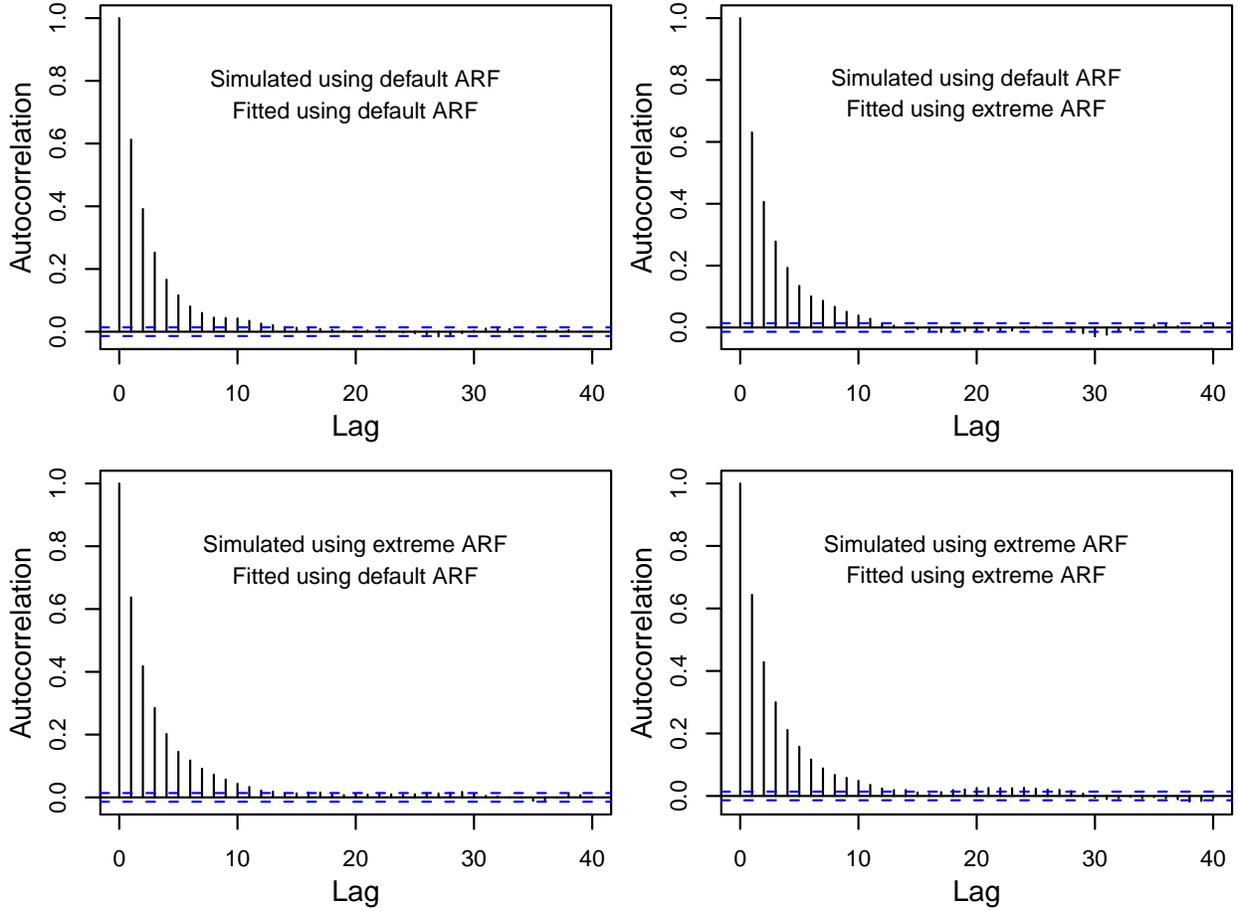}
\caption{
\baselineskip=15pt
The Autocorrelation Function (ACF) of the Parameter Trace in MCMC Runs.  The ACF for the spectral index $\Gamma$ is shown for four cases, where a spectrum is simulated using one effective area curve and the fit is possibly carried out with another.  This explores the dependence of the fitting methodology (codified in the routine {\tt pyBLoCXS}) on misspecified calibration.  The top row shows the ACF for {\sc Simulation~1} (generated using ``default'' effective area curve; see Table~\ref{t:sim}) and the bottom row for {\sc Simulation~5} (generated using an ``extreme'' effective area curve).  The diagonal plots show the ACF when the ``correct'' effective curve is used to fit the spectrum, i.e., the same curve as was used to generate it, and the cross-diagonal plots show the case when the fitting is carried out using a different effective area curve.  The cases in the left column both use the ``default'' effective area to fit the simulated spectra, and the cases in the right column both use the ``extreme'' curve.  The autocorrelation functions demonstrate that $\Gamma^{(k)}$ and $\Gamma^{(k+10)}$ are essentially uncorrelated regardless of whether the correct effective area curve was used in the fit or not. Thus, we set $I=10$ in our pragmatic Bayesian samplers.
}
\label{fig:acf}
\end{center}
\end{figure}

As noted in \S\ref{s:alg:prag}, in order to obtain a sample from the $\theta^{(t)}\sim p(\theta|Y, A^{(t)})$ as in Equation~\ref{eq:mc2} we must iterate {\tt pyBLoCXS} $I$ times to eliminate the dependence of $\theta^{(k-1)}$. To investigate how large $I$ must be, we run {\tt pyBLoCXS} on the spectra from {\sc Simulations~1} and {\sc Simulation~5} of Table~\ref{t:sim}, which were generated using the ``default'' and an ``extreme'' effective area curve. Since {\sc Simulation~5} was generated using the ``extreme'' effective area curve, it is the ``extreme'' curve that is actually ``correct'' and the ``default'' curve that is ``extreme''.  When running {\tt pyBLoCXS} with the ``default'' effective area curve, we initiated the chain at the posterior mean of the parameters given the ``extreme'' curve, and vis versa. This ensures that we are using a relatively extreme starting value and will not underestimate how large $I$ must be to generate an essentially independent draw.  The resulting autocorrelation and time series plots for $\Gamma$ appear in Figure~\ref{fig:acf}.  The autocorrelation plots report the correlation of $\theta^{(k)}$ and $\theta^{(k+I)}$ for each value of $I$ plotted on the horizontal axis. The plots show that for $I>10$ the autocorrelations are essentially zero for both spectra, and we can consider $\theta^{(k)}$ and $\theta^{(k+10)}$ to be essentially independent.  Similarly, the time series plots show that there is no effect of the starting value past the tenth iteration. Similar plots for $N_{\rm H}$ and the normalization parameter (not included) are essentially identical. Thus, in all subsequent computations we set $I=10$ in the pragmatic Bayesian samplers. Generally speaking, the user should construct autocorrelation plots to determine how large $I$ must be in a particular setting.

When we iterate Step~2 in the Pragmatic Bayesian Method, we are more concerned with the mixing of the chain once it has reached its stationary distribution, rather than convergence of the chain to its stationary distribution. This is because convergence to  the stationary distribution will be assessed using the final chain of $\theta^{(t)}$ in the regular way, i.e., using multiple chains (Gelman \& Rubin 1992, van Dyk et al.\ 2001).  Even after the stationary distribution has been reached, we need to obtain a value of $\theta^{(t+1)}$ in Step~2 that is essentially independent of the previous draw, given $A^{(k+1)}$. Thus, we focus on the autocorrelation of the chain $\theta^{(t)}$ for fixed $A$.  This said, if the posterior of $\theta$ is highly dependent on $A$ and $A^{(t)}$ and $A^{(t+1)}$ are extreme within the calibration sample, that the conditional posterior distribution of $\theta$ given $A^{(t)}$ and $A^{(t+1)}$ may be be quite different and we may need to allow $\theta$ to converge to its new conditional posterior distribution. The time series plots in Figure~6 investigate this possibility when extreme values of $A$ are used. Luckily, the effect of these extreme starting values still burns off in just a few iterations, as is evident in Figure~\ref{fig:acf}.

\subsection{Comparing the Algorithms}
\label{s:ex:comp}

We discuss two classes of algorithms in \S\ref{s:alg} to account for calibration uncertainty in spectral analysis: Multiple Imputation, and a pragmatic Bayesian MCMC sampler.  For each, we consider two methods of exploring the calibration product sample space: first by directly sampling from the set of effective areas ${\calA}$, and second by simulating an effective area from a compressed Principal Component representation.  Here, we evaluate the effectiveness of each of the four resulting algorithms, and show that they all produce comparable results, and are a significant improvement over not including the calibration uncertainty in the analysis.  We fit each of the eight simulated data sets described in Table~\ref{t:sim} using each of the four algorithms.  The first four simulations are identical to those described in \S\ref{s:cs:ex}.  Analyses carried out using Multiple Imputation all used $\MI=20$ imputations.  For analyses using the PCA approximation to ${\calA}$, we used $J=17$.  For pragmatic Bayesian methods, we used $I=10$ inner iterations.  Figure~\ref{fig:comp_e5_1to4} gives the resulting estimated marginal posterior distributions for $\Gamma$ for each of the eight simulations and each of the four fitting algorithms along with the results when the effective area is fixed at $A_0$.  Parameter traces (also known as time series) are also shown for all the simulations for the two MCMC algorithms (see \S\ref{s:alg:mc}).  Although the fitted values differ somewhat (see Simulations 1, 2, 3, and 6) among the four algorithms that account for calibration uncertainty, the differences are very small relative to the errors and overall the four methods are in strong agreement. When we do not account for calibration uncertainly, however, the error bars can be much smaller and in some cases the nominal 68\% intervals do not cover the true value of the parameter (see Simulations 1, 2, 5, and 6, corresponding to larger nominal counts). When we do account for calibration uncertainty, only in Simulation~6 did the 68\% intervals not contain the true value, and in this case the 95\% (not depicted) do contain the true value. Results for $\rm N_H$ are similar but omitted from Figure~\ref{fig:comp_e5_1to4} to save space.

An advantage of using MCMC is that it maps out the posterior distribution (under the conditional independence assumptions of Section~\ref{s:meth:stat:marg}) rather than making a Gaussian approximation to the posterior distribution. Notice the non-Gaussian features in the posterior distributions plotted for Simulations 1, 3, 5, and 7 (corresponding to the harder spectral model).

\subsection{Application to a Sample of Radio Loud Quasars}
\label{s:ex:quasar}

Here we illustrate our methods with a realistic case, using X-ray spectra available for a small sample of radio loud quasars observed with the {\it Chandra} X-ray Observatory in 2002 (Siemiginowska et al.\ 2008). We performed the standard data analysis including source extraction and calibration with CIAO software ({\it Chandra} Interactive Analysis of Observations). The X-ray emission in radio loud quasars originates in a close vicinity of a supermassive black hole and could be due to an accretion disk or a relativistic jet.  It is well described by a Compton scattering process and the X-ray spectrum can be modeled by an absorbed power law:
\begin{equation}
S(E)=N~E^{-\Gamma}~e^{-\sigma(E)\,N_{\rm H}}~{\rm~photons~cm^{-2}~sec^{-1}~keV^{-1}} \,, 
\end{equation}
where $\sigma(E)$ is the absorption cross-section, and the three model parameters are: the normalization at 1~keV, $N$; the photon index of the power law, $\Gamma$; and the absorption column, $N_{\rm H}$.

The number of counts in the X-ray spectra varied between 8 and 5500. 
After excluding two datasets (ObsID 3099 which had 8 counts, and ObsID 836 which is better described by a thermal spectrum), we reanalyzed the remaining 15 sources to include calibration uncertainty.
In fitting each source, we included a background spectrum extracted from the same observation over a large annulus surrounding the source region. We adopted a complex background model (a combination of a polynomial and 4 gaussians) that was first fit to the blank-sky data provided by the {\it Chandra} X-ray Center to fix its shape.  While fitting the models to the source and background spectra, we only allow for the normalization of the background model to be free. This is an appropriate approach for very small background counts in the Chandra spectra of point sources. We used this background model for all spectra (except for two -- ObsIDs 3101 and 3106 -- that had short 5~ksec exposure times and small number of counts $<45$, for which the background was ignored). The original analysis (Siemiginowska et al.\ 2008) did not take into account calibration errors, and as we show below the statistical errors are significantly smaller than the calibration errors for sources with a large number of counts.

We fit each spectrum accounting for uncertainty in the effective area in two ways:
\begin{enumerate}
\item with the multiple imputation method in \S\ref{s:alg:mi:pca} using Sherpa for the individual fits, and
\item with the pragmatic Bayesian algorithm in \S\ref{s:alg:prag:pca}  using {\tt pyBLoCXS} for MCMC sampling.
\end{enumerate}
Both of these fits use the PCA approximation using 14 observation-specific default effective area curves, $A_0^*$ in Equation~\ref{eq:pca} with $J=17$. We use $\MI=20$ multiple imputations and $I=10$ subiterations in the pragmatic Bayesian sampler.  To illustrate the effect of accounting for calibration uncertainty, we compared the first fit with the Sherpa fit that fixes the effective area at $A_0^*$ and each of the second and third fits with the {\tt pyBLoCXS} fit that also fixes the effective area at $A_0^*$.

The results appear in Figure~\ref{fig:quasar} which compares the error bars computed with ($\sigma_{\rm tot}$) and without ($\sigma_{\rm stat}$) accounting for calibration uncertainty. The left panel uses \sherpa\ and computes the total error using multiple imputation, and the right panel uses {\tt pyBLoCXS} and computes the total error using the pragmatic Bayesian method.  The plots demonstrate the importance of properly accounting for calibration uncertainty in high-counts, high-quality observations.  The systematic error becomes prominent with high counts because the statistical error is small, and $\sigma_{\rm tot}$ deviates from $\sigma_{\rm stat}$, asymptotically approaching a value of $\sigma_{\rm tot}\approx{0.04}$.  This asymptotic value represents the limiting accuracy of any observation carried out with this instrument, regardless of source strength or exposure duration.  For the absorbed power law model applied here, the systematic uncertainty on $\Gamma$ becomes comparable to the statistical error for spectra with counts $\gtrsim{2400}$, with the largest correction seen in ObsID~866, which had $>14500$ counts.

%NUMERICAL EVALUATION}

%{DISCUSSION

\section{Discussion}
\label{sec:disc} 
In the previous sections, we have worked through a specific example (Chandra effective area) in some detail.  Now, in this section, we present two more complete generalizations.  The first is the case ignored previously, when the data have something interesting to say about the calibration uncertainties.  In the second, we explain how to generalize the algorithms we worked through earlier to the full range of instrument responses, including energy redistribution matrices and point spread functions .

%Finally, we discuss our explicit recommendations for codifying instrument uncertainty in , say, a FITS-file format.

\subsection{A Fully Bayesian Method}
\label{sec:disc:fullbayes}

To avoid the assumption that the observed counts carry little information as to the choice of effective area curve, we can employ a fully Bayesian approach that bases inference on the full posterior distribution $p(\theta, A | Y)$. To do this via MCMC, we must construct a Markov chain with stationary distribution $p(\theta, A | Y)$, which can be accomplished by iterating a two-step Gibbs sampler, for $k=1, \ldots, K$.

\noindent
{\bf A Fully Bayesian Sampler}
\begin{description}
\item[\sc Step~1:] Sample $A^{(k+1)} \sim p(A | \theta^{(k)}, Y)$.
\item[\sc Step~2:] Sample $\theta^{(k+1)}\sim {\cal K}_{\tt pyB} (\theta | \theta^{(k)}; Y ,A^{(k+1)})$.
\end{description}
Notice that unlike in the pragmatic Bayesian approach in \S\ref{s:alg:mc}, {\sc Step~1} of this sampler requires $A$ to be updated given the current data. Unfortunately,  sampling $p(A | \theta^{(k)}, Y)$ is computationally quite challenging. The difficulty arrises because the fitted value of $\theta$ can depend strongly on $A$. That is, calibration uncertainty can have a large effect on the fitted model, see Drake et al.\ (2006) and \S\ref{s:cs:ex}. From a statistical point of view, this means that given $Y$, $\theta$ and $A$ can be highly dependent and $p(A | \theta^{(k)}, Y)$ can depend strongly on $\theta^{(k)}$. Thus a large proportion of the replicates in ${\calA}$ may have negligible probability under $p(A|\theta^{(k)}, Y)$ and it can be difficult to find those that have appreciable probability without doing an exhaustive search. The computational challenges of a fully Bayesian approach are part of the motivation behind our recommendation of the pragmatic Bayesian method. Despite the computational challenges, there is good reason to pursue a Fully Bayesian Sampler. Insofar as the data are informative as to which replicates in ${\calA}$ are more -- or less -- likely, the dependence between $\theta$ and $A$ can help us to eliminate possible values of $\theta$ along with replicates in ${\calA}$, thereby reducing the total error bars for $\theta$. Work to tackle the computational challenges of the fully Bayesian approach is ongoing.

\subsection{General Methods for Handling Calibration Uncertainties}
\label{sec:disc:gen}

In general, the response of a detector to incident photons arriving at time $t$ can be written as
\begin{equation}
M(E^*,\mathbf{x}^*,t;\bfth) =
\int dE \, d\mathbf{x} \
S(E,\mathbf{x},t;\theta) \
R(E,E^*,\mathbf{x}^*;t) \
P(\mathbf{x},\mathbf{x}^*,E;t) \
A(E,\mathbf{x}^*;\mathbf{x},t) \,
\label{eq:resp}
\end{equation}
where $\mathbf{x}^*$ and $E^*$ are the measured photon location and energy (or the detector channel), while $\mathbf{x}$ and $E$ are the true photon sky location and energy; the source physical model $S(E,\mathbf{x},t;\theta)$ describes the energy spectrum, morphology (point, extended, diffuse, etc.), and variability with parameters $\theta$; and $M(E^*,\mathbf{x}^*,t;\theta)$ are the expected counts in detector channel space.  Calibration is carried out using well known instances of $S(E,\mathbf{x},t;\theta)$ to determine the quantities
\begin{eqnarray}
\nonumber
 R(E,E^*,\mathbf{x}^*;t) &\equiv& {\rm Energy \ Redistribution} \\
\nonumber
 P(\mathbf{x},\mathbf{x}^*,E;t) &\equiv& {\rm Point \ Spread \ Function} \\
%\nonumber
 A(E,\mathbf{x}^*;\mathbf{x},t) &\equiv& {\rm  Effective \ Area}
\label{eq:calproducts}
\end{eqnarray}
It is important to note that all of the quantities in Equation~\ref{eq:resp} have uncertainties associated with them.  Our goal is providing a fast, reliable, and robust strategy to incorporate the jittering patterns in all of the calibration products and to draw proper inference, best fits and error bars, reflecting calibration uncertainty.
 
In principle, using a calibration sample to represent uncertainty and the statistical methods for incorporating the calibration sample described in \S\ref{s:meth} and \S\ref{s:alg} can be applied directly to calibration uncertainty for any of the calibration products. The use of PCA, however, to summarize the calibration sample may not be robust enough for higher dimensional and more complex calibration products. More sophisticated image analysis techniques or hierarchically applied PCA may be more appropriate.  Our basic strategy, however, of providing instrument-specific summaries of the variability in the calibration uncertainty and observation-specific measures of the mean (or default) calibration product, is quite general. Thus, in this section, we focus on the generalization of Equation~\ref{eq:pca0} and begin by rephrasing the equation as
%\begin{equation}
%{\rm Simulated} = {\rm Nominal} + {\rm Bias} + {\rm random~components} + {\rm residuals} \,.
%\end{equation}
%
\begin{equation}
{\rm Replicate~Calibration~Product} = {\rm Mean} + {\rm Offset} + {\rm Explained~Variability} + {\rm Residual~Variability} \,.
\label{eq:gencompgen}
\end{equation}
Here the Mean is the mean of the calibration sample, the Offset is the shift that we impose on the center of distribution of the calibration uncertainty to account for observation-specific differences, the Explained Variability is the portion of the variability that summarize in parametric and/or systematic way (e.g., using PCA), and the Residual Variability is the portion of the variability left unexplained by the systematic summary. These four terms correspond to the four terms in Equation~\ref{eq:pca0}.
%Because omitting the offset term will lead to {\it bias} in the fitted values, one might also call this the ``Bias''.

The formulation in Equation~\ref{eq:gencompgen} removes the necessity of depending solely on PCA to summarize variance in the calibration sample, and allows us to use a variety of methods to generate the simulated calibration products.  For example, we can even include such loosely stated measures of uncertainty as ``the effective area is uncertain by X\% at wavelength Y''. This formulation is not limited to describing effective areas alone, but can also be used to encompass the calibration uncertainty in response matrices and point spread functions.  The precise method by which the variance terms are generated may vary widely, but in all foreseeable cases they can be described as in Equation~\ref{eq:gencompgen}, with an offset term and a random variance component added to the mean calibration product, and with an optional residual component. The calibration sample simulated in this way form an informative prior $p(A,R,P)$ that could be used like $p(A)$ in Equation~\ref{eq:mc1}.  Some potential methods of describing the variance terms are:

\begin{enumerate}
\item
When a large calibration sample is available, the random component is simply the full set of calibration products in the sample. When using a Monte Carlo for model fitting, as in \S\ref{s:meth:stat}.3,  a random index is chosen at each iteration and the calibration product  corresponding to that index is used for that iteration.  This process preserves the weights of the initial calibration sample.  In this scenario the residual component is identically zero.

\item
If the calibration uncertainty is characterized by a multiplicative polynomial term in the source model, the explained variance component in Equation~\ref{eq:gencompgen} can be obtained by sampling the parameters of the polynomial, from a Gaussian distribution, using their best-fit values and the estimated errors. These simulated calibration products can then be used to modify the nominal products inside each iteration.  Thus, the offset and residual terms are zero, and only the polynomial parameter best-fit values and errors need to be stored.

\item
If a calibration product is newly identified, it may be systematically off by a fixed but unknown amount over a small passband, and users can specify their own estimate of calibration uncertainty as a randomized additive constant term over the relevant range.  This is essentially equivalent to using a correction with a first-order polynomial. The stored quantities are the average offset, the bounds over which the offset can range, and a pointer specifying whether to generate uniform or Gaussian deviates over that range.
\end{enumerate}

%DISCUSSION}

%{SUMMARY
\section{Summary}
\label{sec:summ}

We have developed a method to handle in a practical way the effect of uncertainties in instrument response on astrophysical modeling, with specific application to \chandra/ACIS instrument effective area. Our goal has been to obtain realistic error bars on astrophysical source model parameters that include both statistical and systematic errors. For this purpose, we have developed a general and comprehensive strategy to describe and store calibration uncertainty and to incorporate them into data analysis. Starting from the full, precise, but cumbersome objective-function of the parameters, data, and instrument uncertainties, we adopt a Bayesian posterior-probability framework and simplify it in a few key places to make the problem tractable. This work holds practical promise for a generalized treatment of instrumental uncertainties in not just spectra but also imaging, or any kind of higher-dimensional analyses; and not just X-rays, but across wavelengths and even to particle detectors. Our scheme treats the possible variations in calibration as an informative prior distribution while estimating the posterior probability distributions of the source model parameters. Thus, the effects of calibration uncertainty is automatically included in the result of a single fit. This is different from a usual sensitivity study in that we provide an actual uncertainty estimate.  Our analysis shows that systematic error contribution in high counts spectra is more significant than when there are few counts; therefore, including calibration uncertainty in a spectral fitting strategy is highly recommended for high quality data.

We adopt the calibration uncertainty variations, in particular the effective area variations for the {\sl Chandra}/ACIS-S detector, described by Drake et al.\ (2006), as an exemplar case.  Using the effective area sample $\calA$ simulated by them, we 
\begin{enumerate}
\item show that variations in effective areas lead to large variations in fitted parameter values;
\item demonstrate that systematic errors are relatively more important for high counts, when statistical errors are small;
\item describe how the calibration sample can be effectively compressed and summarized by a small number of components from a Principal Components Analysis;
\item outline two separate algorithms with which to incorporate systematic uncertainties within spectral analysis:
\begin{enumerate}
\item an approximate, but quick method based on the Multiple Imputation combining rule that carries out spectral fits for different instances of the effective area and merges the mean of the variances with the variance of the means; and
\item a pragmatic Bayesian method that incorporates sampling of the effective areas as from a prior distribution within an MCMC iteration scheme.
\end{enumerate}
\item detail two methods of sampling $A\rep$: directly from the calibration sample $\calA$, and via a PCA decomposition
\item show that $\approx{20}$ representative samples of $A\rep$ are needed to obtain relatively reliable estimates of uncertainty;
\item apply the method to a real dataset of a sample of quasars and show that known systematic uncertainties require that, e.g., the power-law index $\Gamma$ cannot be determined with an accuracy better than $\sigma_{\rm tot}(\Gamma)\approx{0.04}$; and
\item discuss future directions of our work, both in relaxing the constraint of not allowing the calibration sample $\calA$ to be affected by the data, and in generalizing the technique to other sources of calibration uncertainty.
\end{enumerate}
%We anticipate that this mechanism ({\tt pyBLoCXS}) of incorporating calibration uncertainties will become available via {\sl CIAO}/\sherpa.

\acknowledgments
This work was supported by NASA AISRP grant NNG06GF17G (HL, AC, VLK), and CXC NASA contract NAS8-39073 (VLK, AS, JJD, PR), NSF grants DMS 04-06085 and DMS 09-07522 (DvD, AC, SM, TP), and NSF grants DMS-0405953 and DMS-0907185 (XLM). We acknowledge useful discussions with Herman Marshall, Alex Blocker, Jonathan McDowell, and Arnold Rots.

%SUMMARY}

\newpage

%{BIBLIOGRAPHY

\newpage

%BIBLIOGRAPHY}

\newpage
%{FIGS AND TABLES

\addtocounter{figure}{-1}
\begin{figure}[t]
\begin{center}
\includegraphics[width=6.5in,angle=0]{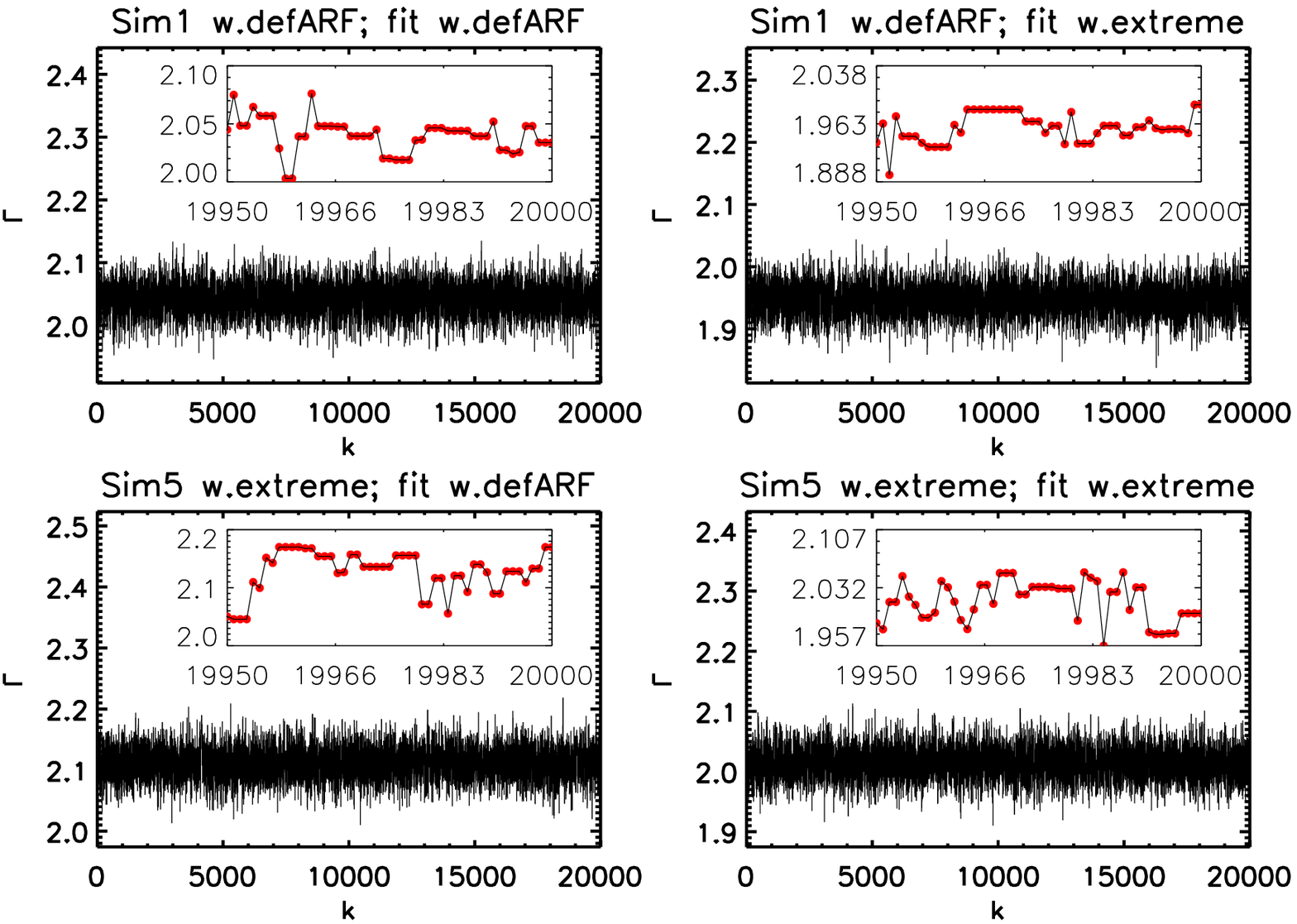}
\caption{
\baselineskip=15pt
{\sc (contd.)} The parameter traces for the spectral index $\Gamma$, shown for same cases as the autocorrelation cases shown before.
While the autocorrelation determines the ``stickiness'' of the MCMC iterations, the time series demonstrates that choosing misspecified calibration files does not have any effect on the convergence of the solutions.
The traces are shown in the same order as before, for all iterations $k$.
The inset shows the last 50 iterations, with $\Gamma^{(k)}$ denoted by filled circles, and consecutive iterations connected by thin straight lines.
The necessity of using $I>>1$ is apparent in the slow changes in the values of $\Gamma^{(k)}$.
}
\label{fig:trace_acf}
\end{center}
\end{figure}

\begin{figure}[t]
\begin{center}
\includegraphics[width=3.2in]{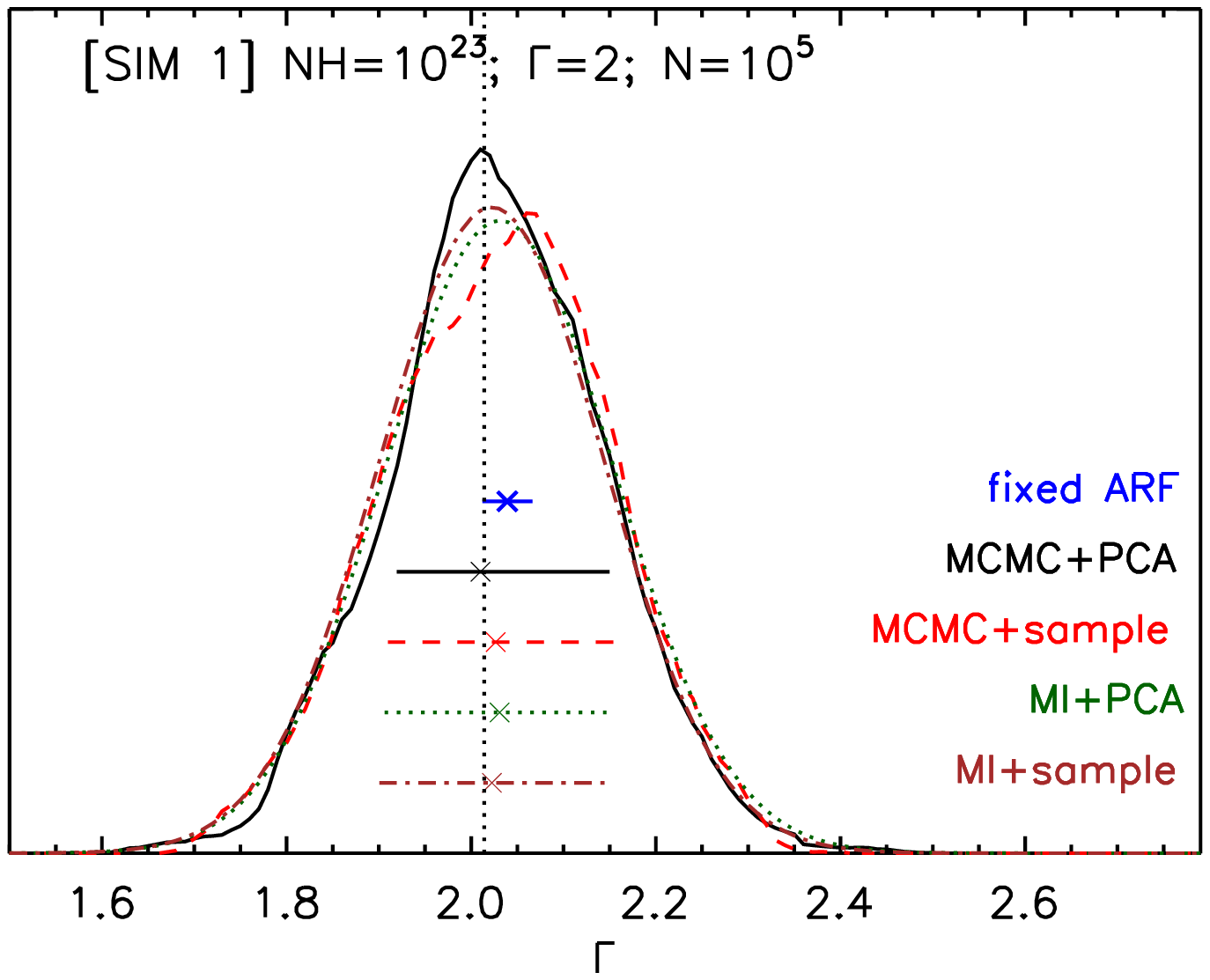}
\includegraphics[width=3.2in]{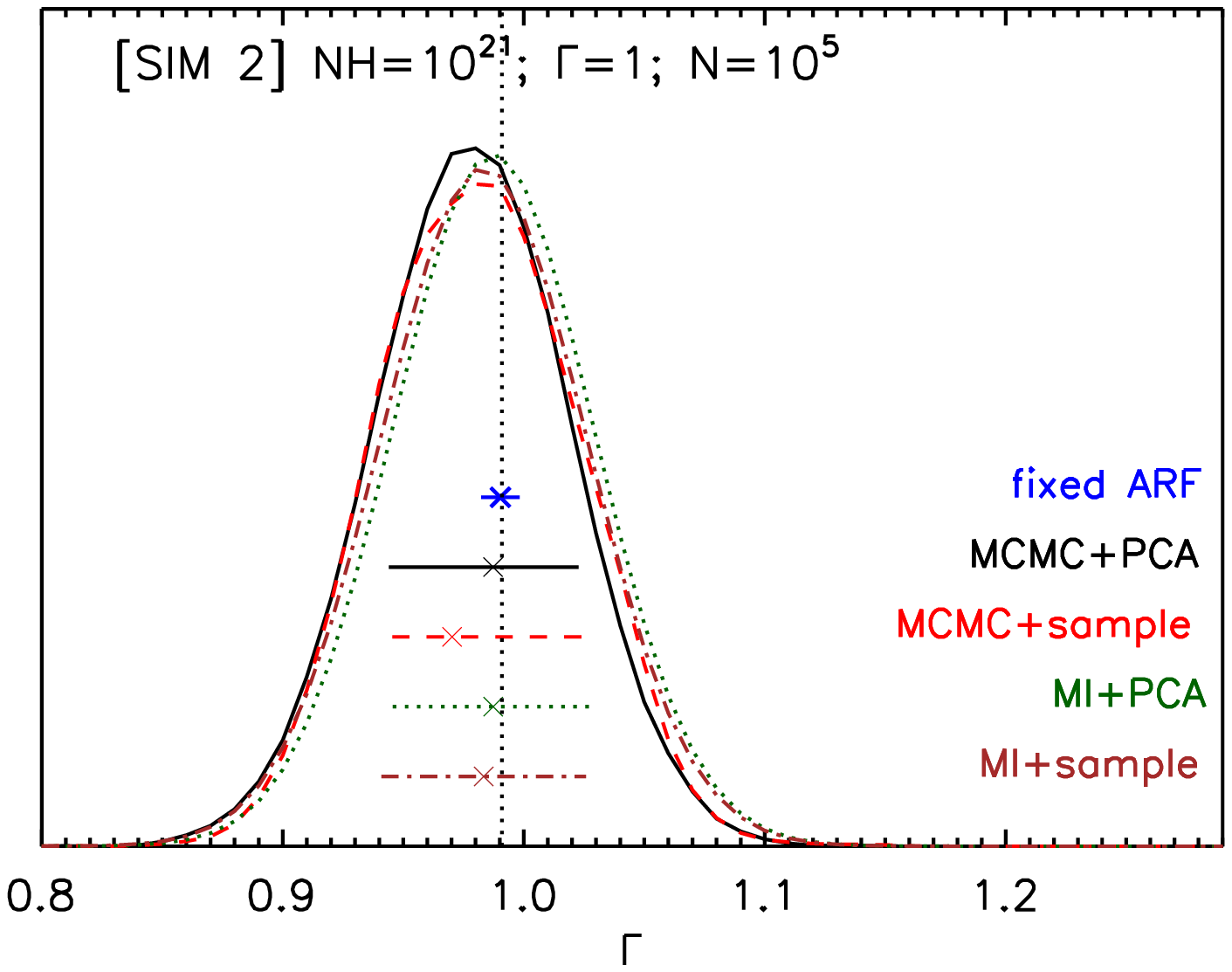} \\
\includegraphics[width=3.2in]{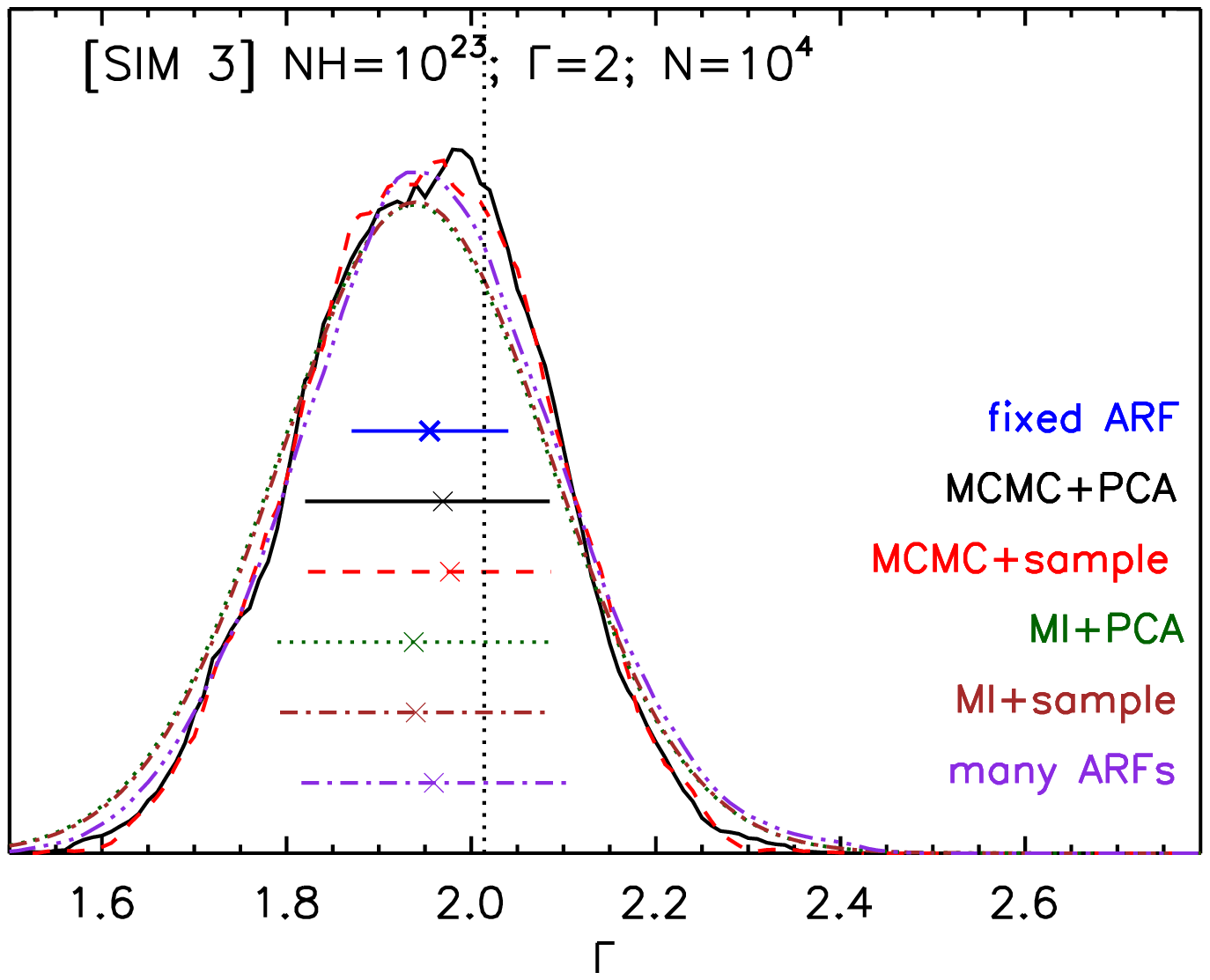}
\includegraphics[width=3.2in]{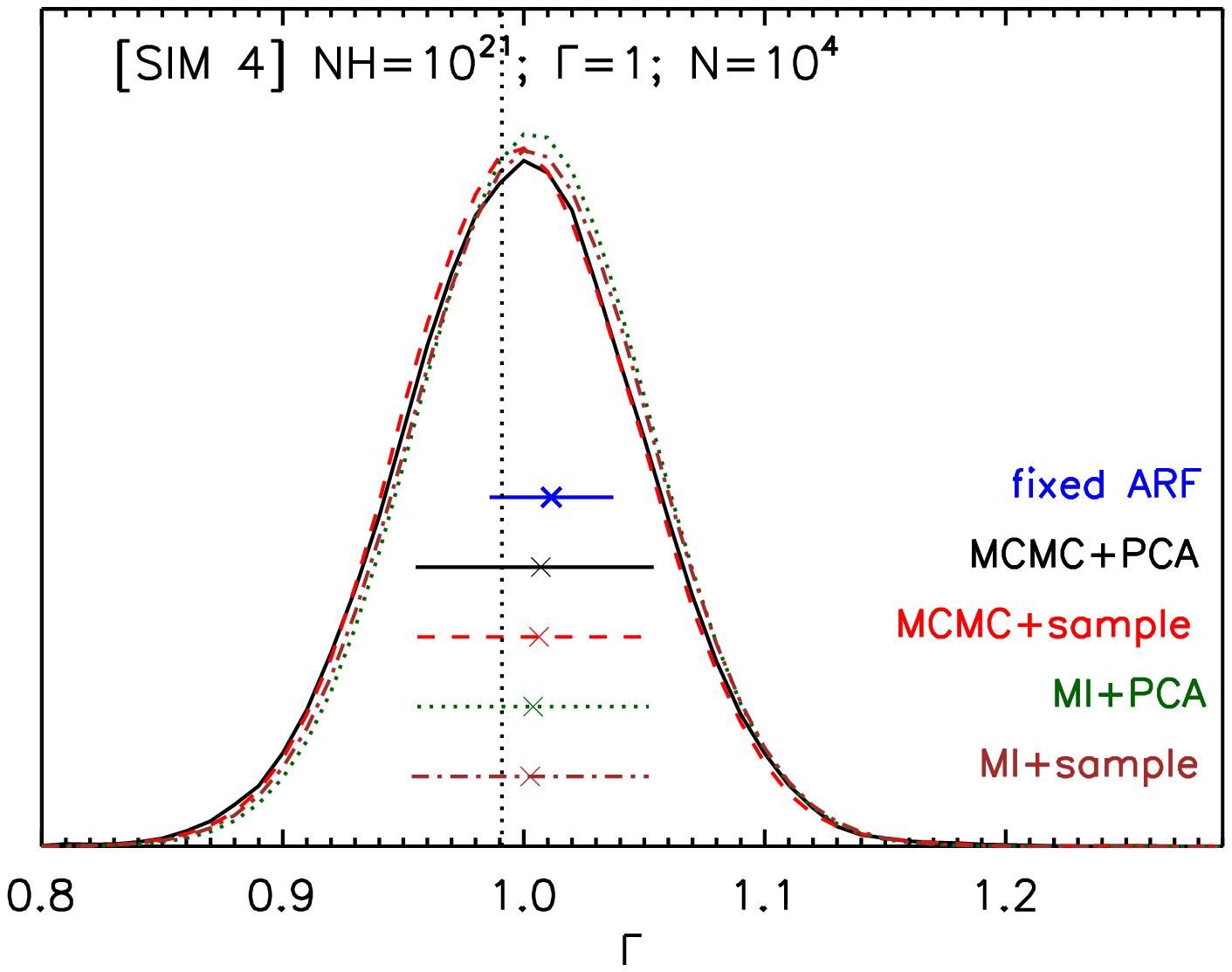}
\caption{
\baselineskip=15pt
Comparing the Algorithms in \S\ref{s:ex} as Applied to the Simulated Spectra 1-4 in Table~\ref{t:sim}.  These are spectra which are generated using the default effective area.  The ``true'' value of the power-law index parameter that was used to generate the simulated spectra is shown as the vertical dashed line.  For each simulation, posterior probability density functions of the power-law index parameter are computed using the pragmatic Bayesian with PCA (black solid curve; \S\ref{s:alg:mc}.4), pragmatic Bayesian with sampling from ${\calA}$ (red dashed curve; \S\ref{s:alg:mc}.3), Multiple Imputation with PCA (green dotted curve; \S\ref{s:alg:mi}.2), Multiple Imputation with samples from ${\calA}$ (brown dot-dashed curve; \S\ref{s:alg:mi}.1), and the combined posteriors from individual runs using the full sample ${\calA}$ (purple dash-dotted curve).  Results for the column density parameter $N_{\rm H}$ are similar.  We use $\MI=20$ samples for multiple imputation.  The density curves are obtained from smoothed histograms of MCMC traces from {\tt pyBLoCXS} for the Bayesian cases, and are Gaussians with the appropriate mean and variance obtained via fitting with {\tt XSPEC\,v12} for the Multiple Imputation cases.  Also shown are the 68\% equal-tail intervals as horizontal bars, with the most probable value of the photon index indicated with an `x' for each of these case, and additionally for the case where a fixed effective area was used to obtain only the statistical error.  Note that in all cases, fitting with the default effective area alone leads to an underestimate of the true uncertainty in the fitted parameter.
}
\label{fig:comp_e5_1to4}
\end{center}
\end{figure}

\addtocounter{figure}{-1}
\begin{figure}[t]
\begin{center}
\includegraphics[width=3.2in]{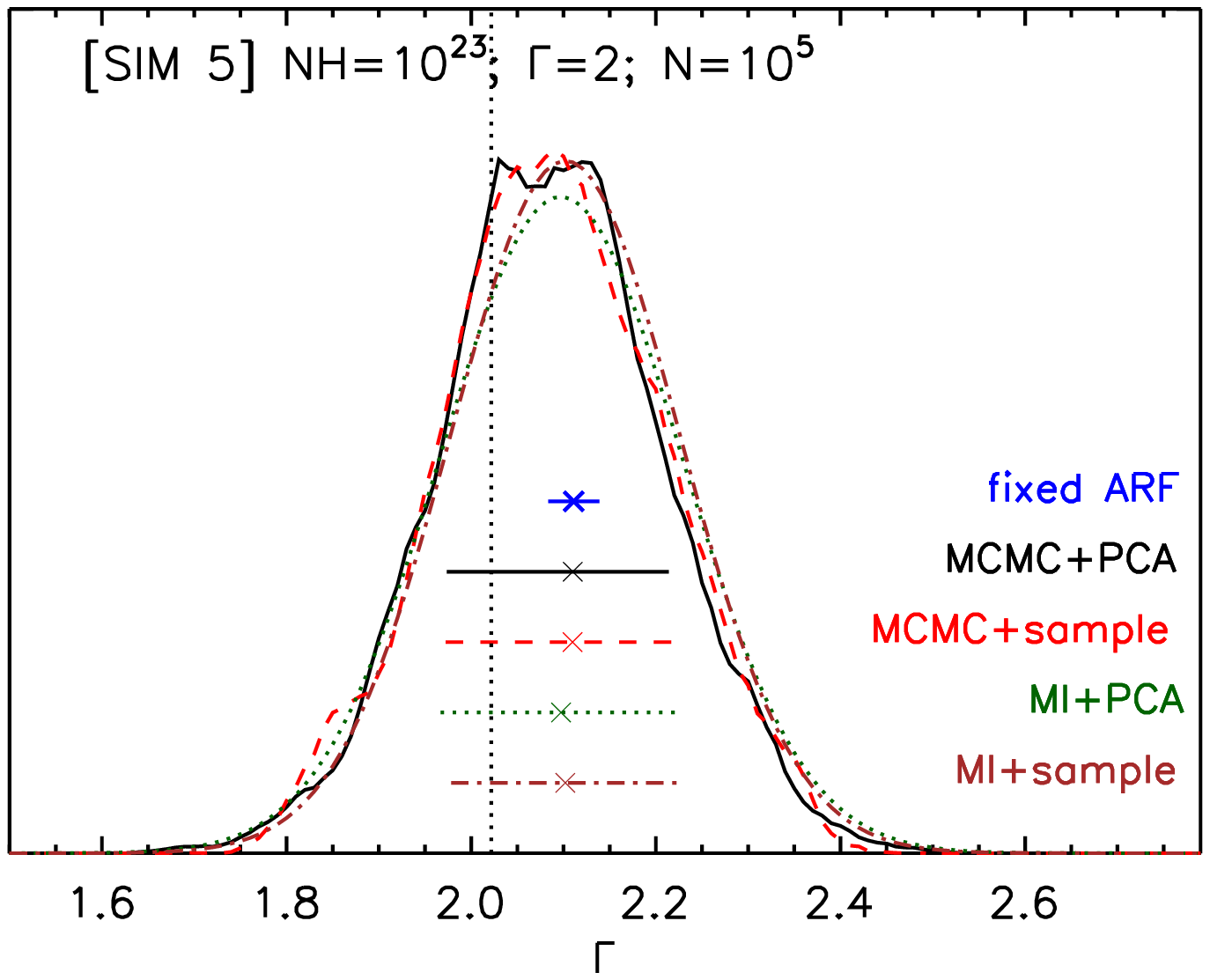}
\includegraphics[width=3.2in]{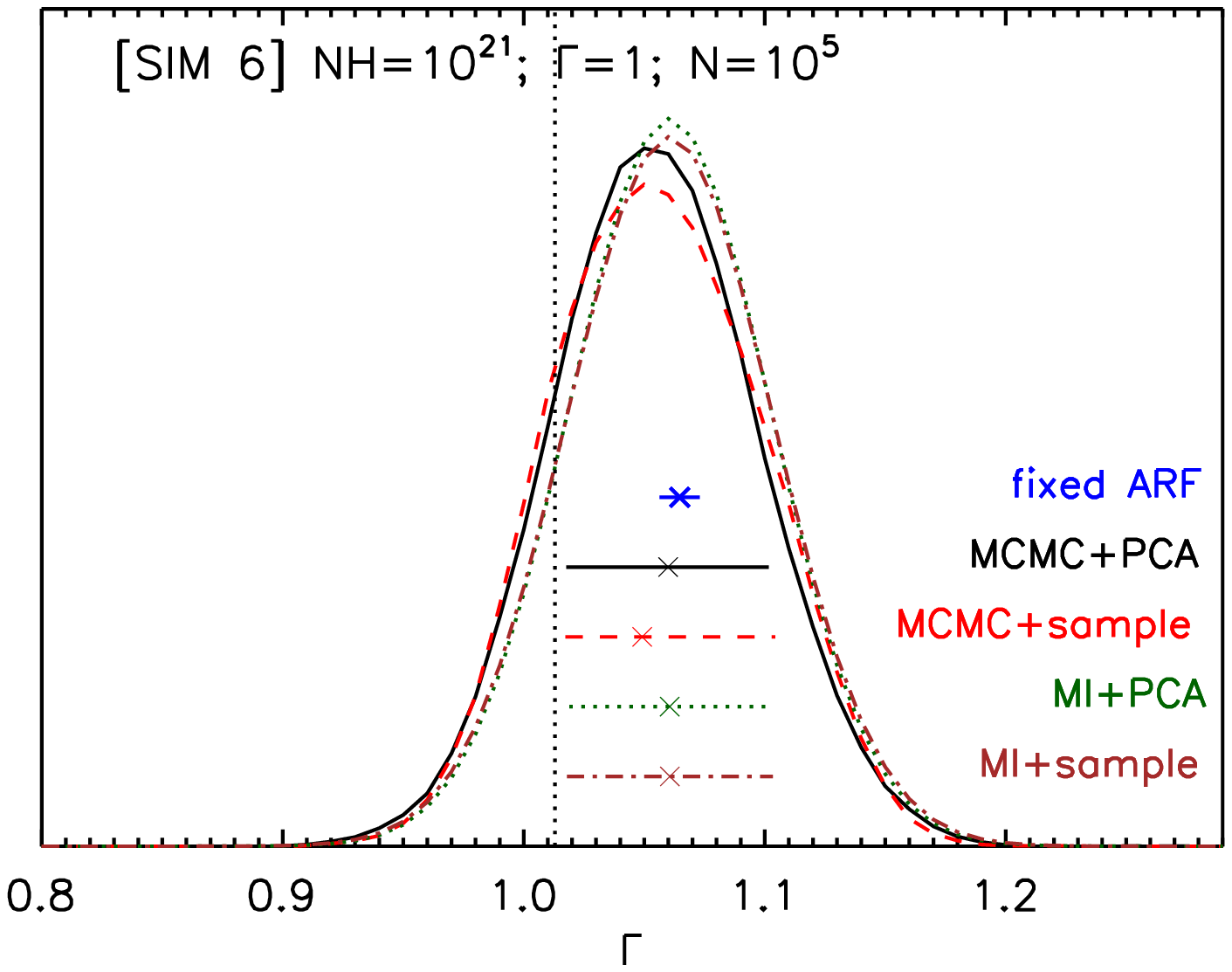} \\
\includegraphics[width=3.2in]{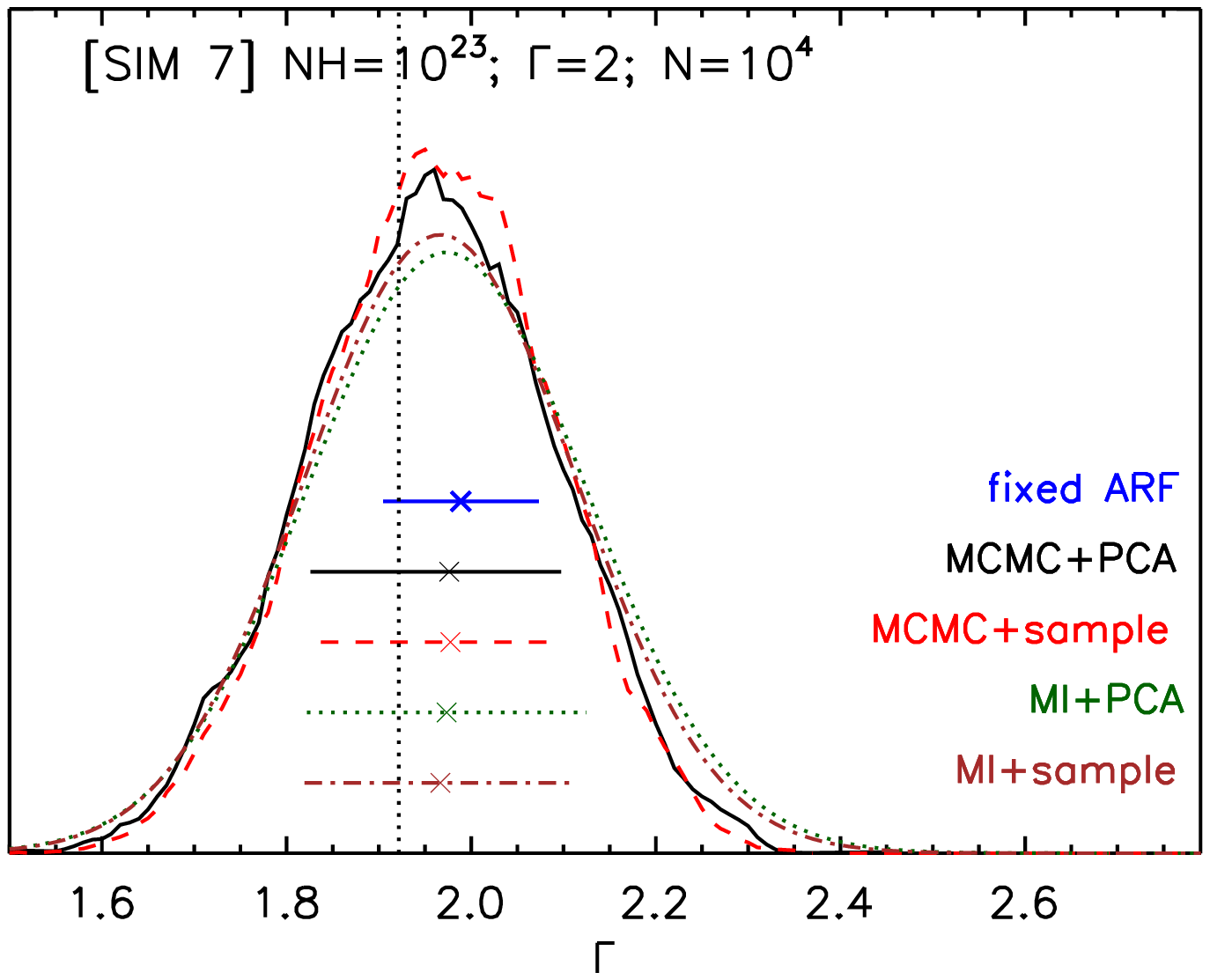}
\includegraphics[width=3.2in]{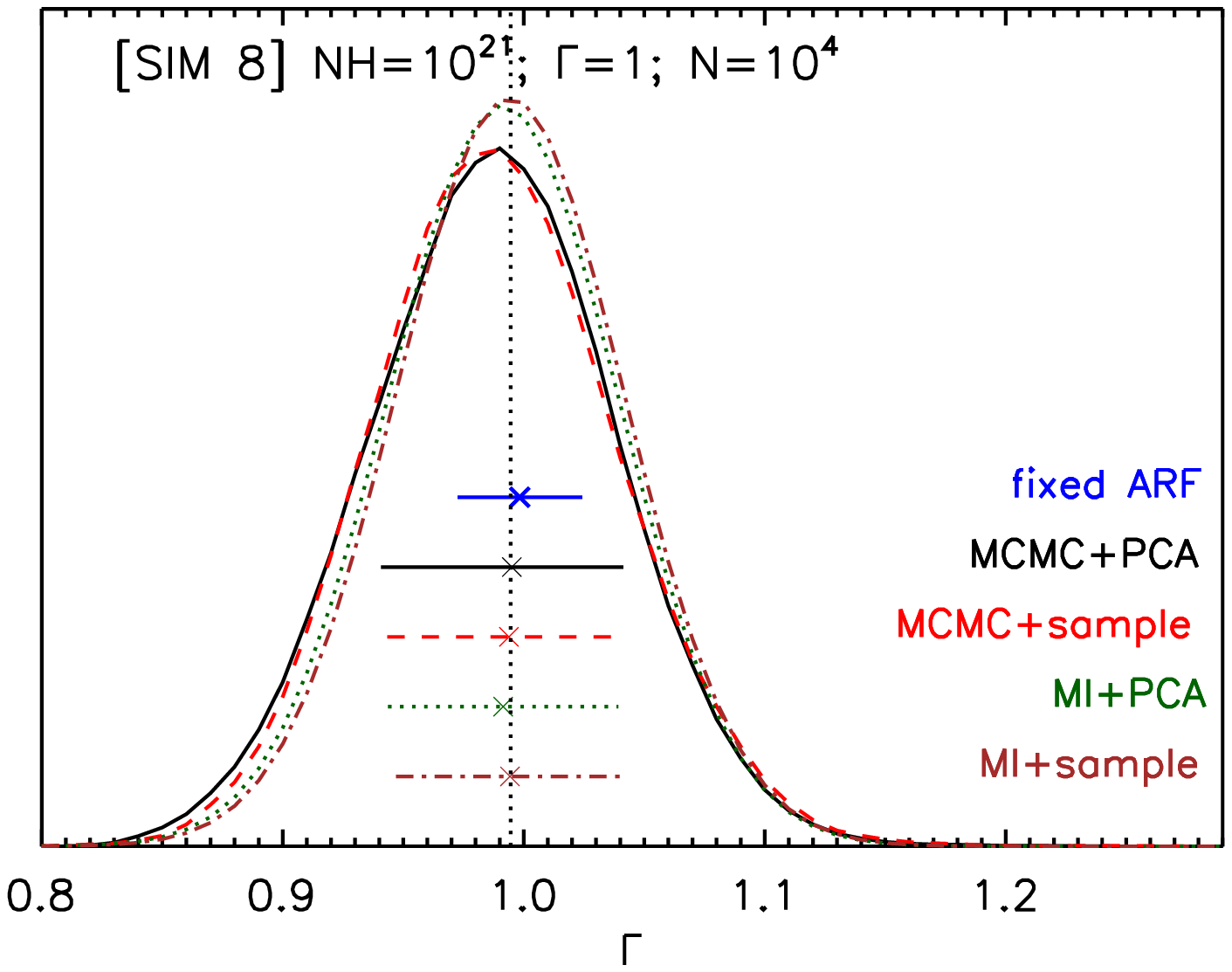}
\caption{
\baselineskip=15pt
{\sc (contd.)} For Simulated Spectra 5-8 in Table~\ref{t:sim}.  These are spectra which are generated using an extreme instance of an effective area from ${\calA}$.  The fits when only one effective area is used are done with the default effective area.  Note that in many cases, not incorporating the calibration uncertainties results in intervals for the parameter which does not contain the true value.
}
\label{fig:comp_e5_5to8}
\end{center}
\end{figure}

\addtocounter{figure}{-1}
\begin{figure}[t]
\begin{center}
\includegraphics[width=6.8in]{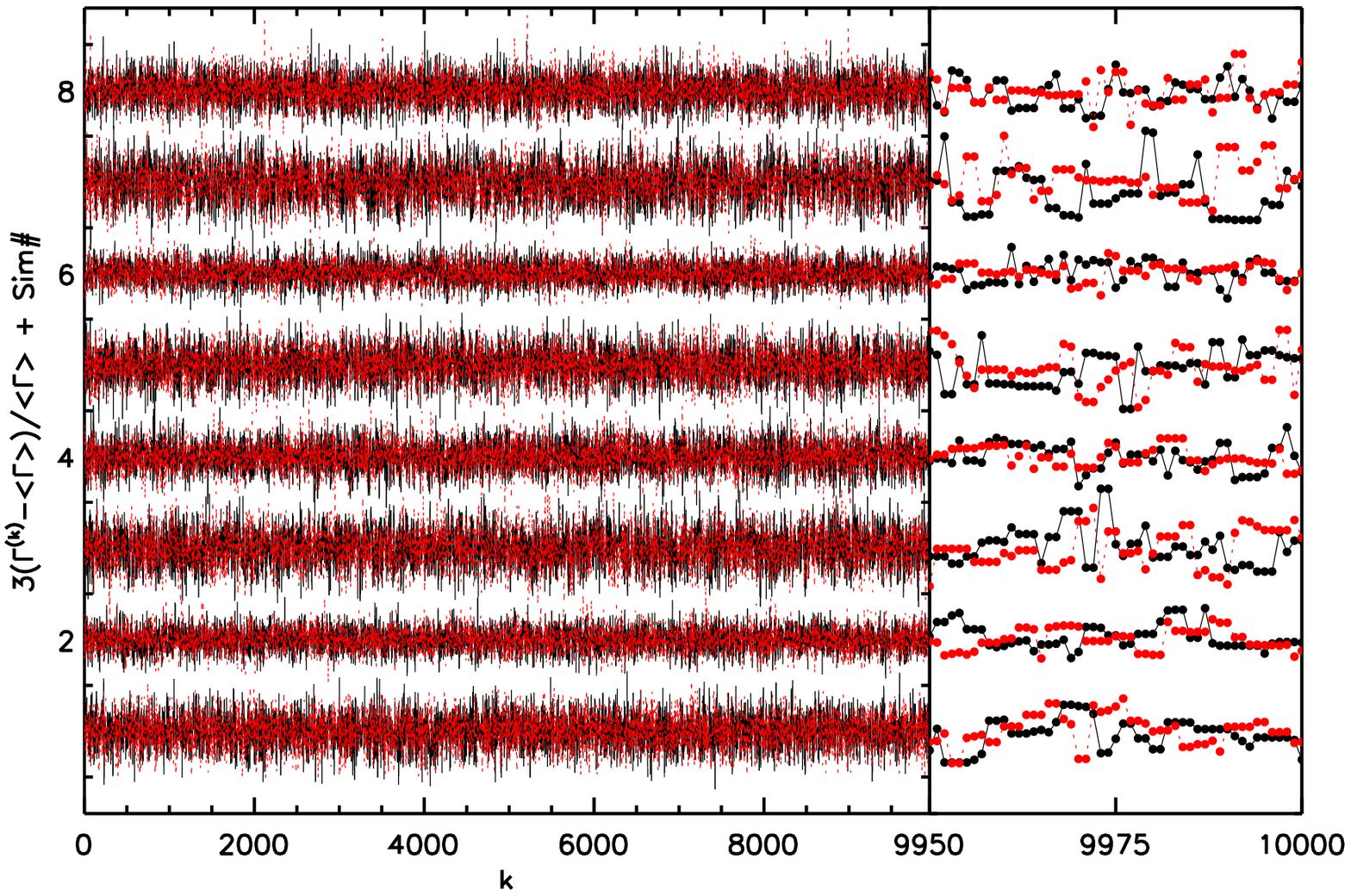}
\caption{
\baselineskip=15pt
{\sc (contd.)} Parameter traces for the spectral index $\Gamma$ for each of the 8 simulations.
All the simulations are shown on the same plot, rescaled (to depict the fractional deviation from the mean, inflated by a factor of 3) and offset (by an integer corresponding to the number assigned to the simulation) for clarity.
The traces for both the {\tt MCMC+PCA} (pragmatic Bayesian algorithm using PCA to generate new effective areas; solid black lines) and {\tt MCMC+sample} (pragmatic Bayesian algorithm with sampling from ${\calA}$; dotted red lines) are shown, with the latter overlaid on the former.
The last 50 iterations are shown zoomed out in the absissa for clarity, and shows each transformed $\Gamma^{(k)}$ as filled circles, connected by thin lines of the corresponding style and color.
Note that all iterations $k$ are shown, but in the calculations of the posterior probability distributions, only every $I^{th}$ iteration, where $I=10$, is used (see Figure~\ref{fig:acf}).
}
\label{fig:comp_e5_traces}
\end{center}
\end{figure}

\begin{figure}[t]
\begin{center}
\includegraphics[width=3.2in]{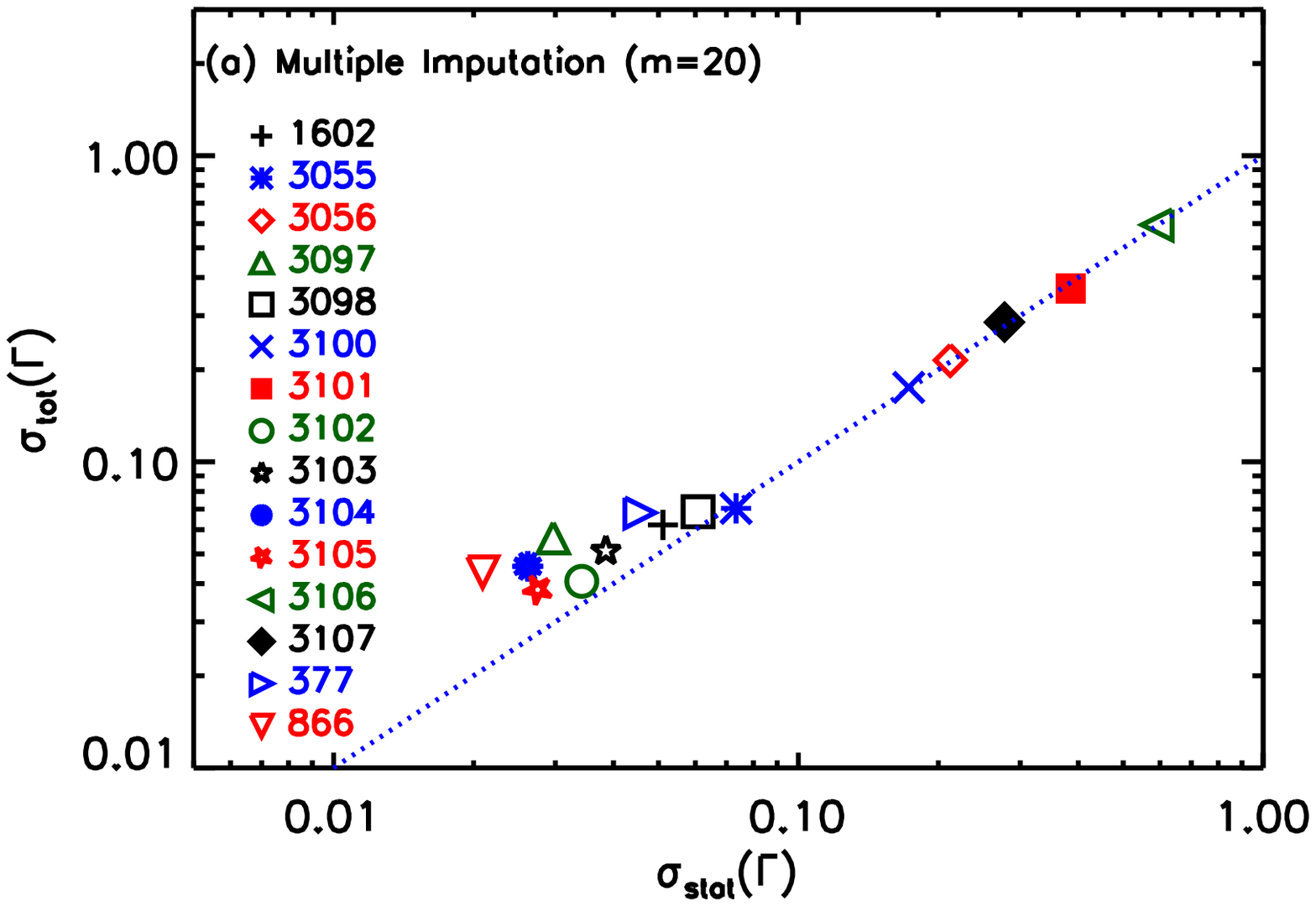}
\includegraphics[width=3.2in]{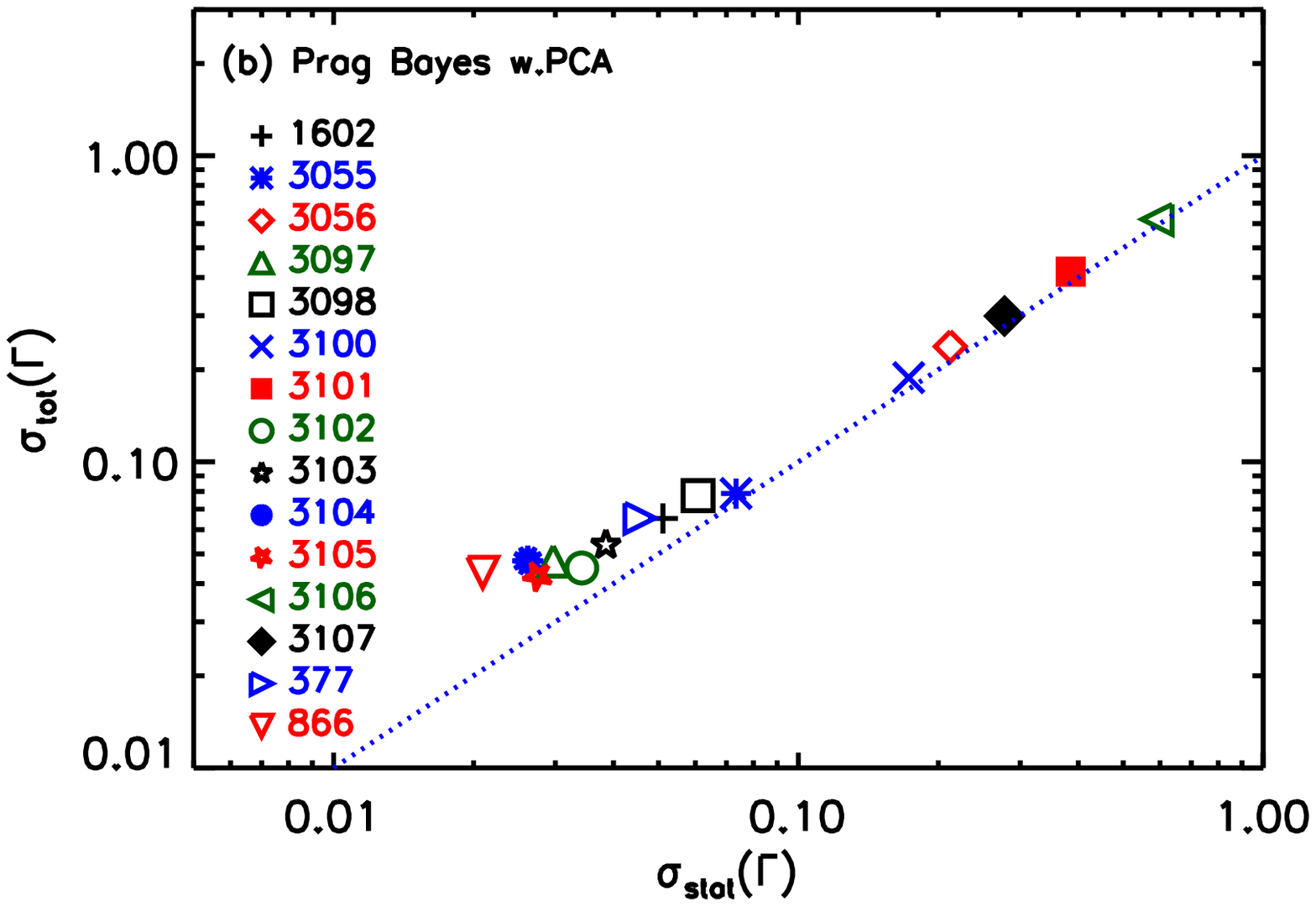}
\caption{
Comparison of the Statistical Error with the Total Error Including Effective Area Uncertainties for Different Methods of Evaluating Them.  Results of fits to a sample of 15 radio loud quasars (Siemiginowska et al.\ 2008; see \S\ref{s:ex:quasar}) are shown.  The abscissae represent the statistical uncertainty $\sigma_{\rm stat}$ as derived by adopting a fixed, nominal effective area, and fit with absorbed power-law models using {\sl CIAO}/Sherpa (stronger sources tend to have smaller error bars).  They are compared with the total error, $\sigma_{\rm tot}$ derived using (a) the Multiple Imputation combining rule (\S\ref{s:alg:mi}.2) with {\sl CIAO}/Sherpa ($M=20$), and (b) the pragmatic Bayesian method with PCA (\S\ref{s:alg:mc}.4), with {\tt pyBLoCXS}.  (Similar results are obtained when using the pragmatic Bayesian method for the full sample of effective areas.)  The different symbols correspond to the analysis carried out for different observations.  The dotted line represents equality, where the total error is identical to the statistical error.  The systematic error cannot be ignored when the statistical error is small, and represents the limiting accuracy of a measurement.
}
\label{fig:quasar}
\end{center}
\end{figure}

\begin{deluxetable}{c|l}
\tablecolumns{2}
\tablecaption{Glossary of symbols used in the text\label{tab:glossary}}
\tablehead{\colhead{Symbol} & \colhead{Description}}
\startdata
%\begin{table}[ht]
%\caption{Glossary of symbols used in the text}\label{tab:glossary}
%\begin{tabular}{c|l}
%\hline
$A$ & effective area (ARF) curve \\
$A^{\rep}$ & replicate $A$ generated from PCA representation of the calibration sample\\
$A_0$ & the default effective area curve.\\
$A_0^*$ & the observation specific effective area curve.\\
$A_l$ & effective area curve $l$ in the calibration sample \\
$\calA$ & a set of effective areas, the calibration sample \\
$\delta \bar A$ & average offset of $\calA$ from $A_0$ \\
$B$ & the between imputation (or systematic) variance of $\hat\theta$. \\
$B_{mm}$ & diagonal element $m$ of $B$\\
$E$ & energy of incident photon \\
$E^*$ & energy channel at which the detector registers the incident photon \\
$e_j$ & random variate generated from the standard Normal distribution \\
$f_l$ & fractional variance of component $l$ in the PCA representation \\
$I$ & number of inner iterations in {\tt pyBLoCXS}, typically $10$ \\
$J$ & number of components used in PCA analysis, here $17$ \\
$j$ & principal component number or index \\
$^{(k)}$ & the superscript indicates the running index of random draws \\
${\cal K}$ & an MCMC kernel \\
${\cal K}_{\tt pyB}$ & the MCMC kernel used in {\tt PyBLoCKS} \\
$L$ & number of replicate effective area curves in calibration sample\\
$l$ & replicate effective area number or index, or principal component number \\
$m$ & imputation number or index \\
$\MI$ & number of imputations \\
${\cal M}$ & response of a detector to incident photons, see Equation~\ref{eq:sim_arf} \\
$p$ & objective function (posterior distribuiton, likelihood, or perhaps $\chi^2$)\\
$P$ & point spread function (PSF) \\
$R$ & energy redistribution matrix (RMF) \\
$r_l^2$ & eigenvalue or PC coefficient of component $l$ in the PCA representation \\
$S$ & astrophysical source model \\
$T$ & total variance of $\hat\theta$.\\
$v_l$ & eigen- or feature-vector for component $l$ in the PCA representation \\
$W$ & the within imputation (or statistical) variance of $\hat\theta$.\\
$W_{mm}$ & diagonal elements $m$ of $W$\\
${\bf x}$ & true sky location of photons \\
${\bf x}^*$ & locations of incident photons as registered by detector \\
$Y$ & data, typically used here as counts spectra in detector PI bins \\
$Z$ & data and physical calculations used by calibration scientists \\ 
$\theta$ & model parameter of interest \\
$\hat\theta$ & estimate of $\theta$ \\
$\hat\theta_m$ & estimate of $\theta$ corresponding to imputed effective area $m$ \\
${\rm Var}(\hat\theta_m)$ & estimates variance of $\hat\theta_m$ \\
$\sigma_{\rm stat}$ & $\sqrt{W}$, representing the statistical error on $\theta$ \\
$\sigma_{\rm tot}$ & $\sqrt{T}$, representing the total error on $\theta$ \\
$\xi$ & a sum of the smaller components, J+1 to L in the PCA representation \\
%\hline
%\end{tabular}
%\end{table}
\enddata
\end{deluxetable}

\begin{table}[t]
\caption{The Eight Simulations Used to Compare the Four Algorithms Described in \S\ref{s:alg}.}
\label{t:sim}
\medskip

\begin{tabular}{lccccccccc}
\hline
& &  \multicolumn{2}{c}{Effective Area}& & \multicolumn{2}{c}{Nominal Counts} & &\multicolumn{2}{c}{Spectal Model}  \\
\cline{3-4}
\cline{6-7}
\cline{9-10}
& & Default & Extreme && \quad$10^5$ & \quad$10^4$ && Hard$^\dagger$ & Soft$^\ddagger$ \\
\hline
{\sc Simulation~1} && X &   && X &   && X &   \\
{\sc Simulation~2} && X &   && X &   &&   & X \\
{\sc Simulation~3} && X &   &&   & X && X &   \\
{\sc Simulation~4} && X &   &&   & X &&   & X \\
{\sc Simulation~5} &&   & X && X &   && X &   \\
{\sc Simulation~6} &&   & X && X &   &&   & X \\
{\sc Simulation~7} &&   & X &&   & X && X &   \\
{\sc Simulation~8} &&   & X &&   & X &&   & X \\
\hline
\end{tabular}

$^\dagger${\footnotesize An absorbed powerlaw with $\Gamma=2$, $N_{\rm H}=10^{23}/{\rm cm}^2$} \\
$^\ddagger${\footnotesize An absorbed powerlaw with $\Gamma=1$, $N_{\rm H}=10^{21}/{\rm cm}^2$}
\end{table}

%FIGS AND TABLES}

\end{document}